\documentclass[aps, twocolumn, superscriptaddress]{revtex4-1}

\usepackage{amsfonts}           % AMS packages
\usepackage{amssymb}            % ^
\usepackage{amsmath}            % ^
\usepackage{comment}
\usepackage{standalone}
\usepackage[utf8]{inputenc}
\usepackage{latexsym}           % For the "normal" symbol (and other funky binary operators)
\usepackage{dcolumn}            % Align table columns on decimal point
\usepackage{graphicx}   % For production with eps / jpg figures
\usepackage{enumerate}
\usepackage{epstopdf}           % Can incorporate eps in the above pdftex
\usepackage{fancyhdr}           % Does exactly what it says it will
\usepackage{hyperref}           % Allow for hyperlinks
\usepackage{setspace}           % Allows us to change spacings
\usepackage{xspace}             % To include spaces after newly defined commands when they are expecting arguments
\usepackage{subfigure}          % Gives us the ability to put several figures side-by-side within one.
\usepackage{bm}                 % A better "bold math" package -- any math symbol can be emboldened with the command \bm{*}
\usepackage[ulem=normalem]{changes}
\usepackage{url}
\usepackage{color}
\newcommand{\bea}{\begin{eqnarray}}
\newcommand{\eea}{\end{eqnarray}}

\newcolumntype{L}[1]{>{\raggedright\let\newline\\\arraybackslash\hspace{0pt}}m{#1}}
\newcolumntype{C}[1]{>{\centering\let\newline\\\arraybackslash\hspace{0pt}}m{#1}}
\newcolumntype{R}[1]{>{\raggedleft\let\newline\\\arraybackslash\hspace{0pt}}m{#1}}

\begin{document}

%Title.  Two "affiliation" class options are groupedaddress (default) and superscriptaddress
\title{The Emergence and Role of Dipolar Dislocation Patterns in Discrete and Continuum Formulations of Plasticity}
\date{\today}

\author{P\'eter Dus\'an Isp\'anovity}
\email{ispanovity@metal.elte.hu}
\affiliation{Department of Materials Physics, E\"otv\"os University, P\'azm\'any P\'eter s\'et\'any 1/a, H-1117 Budapest, Hungary}

\author{Stefanos Papanikolaou}
\affiliation{Department of Mechanical Engineering, The West Virginia University}
\affiliation{Department of Physics, The West Virginia University}
\affiliation{Department of Mechanical Engineering, The Johns Hopkins University,  Baltimore, MD 21218}

\author{Istv\'an Groma}
\affiliation{Department of Materials Physics, E\"otv\"os University, P\'azm\'any P\'eter s\'et\'any 1/a, H-1117 Budapest, Hungary}

\begin{abstract}
{  The plasticity transition at the yield strength of a crystal typically signifies the tendency of dislocation defects towards relatively unrestricted motion. For an isolated dislocation the motion is in the slip plane with velocity proportional to the shear stress, while due to the long range interaction dislocation ensembles move towards satisfying emergent collective elastoplastic modes. Such collective motions have been discussed in terms of the elusively defined {\it backstress}. In this paper, we develop a two-dimensional stochastic continuum dislocation dynamics theory that clarifies the role of backstress and demonstrates a precise agreement with the collective behavior of its discrete counterpart, as a function of applied load and with only three essential free parameters. The main ingredients of the continuum theory is the evolution equations of statistically stored and geometrically necessary dislocation densities, which are driven by the long-range internal stress, a stochastic flow stress term and, finally, two local ``diffusion'' like terms. The agreement is shown primarily in terms of the patterning characteristics that include the formation of dipolar dislocation walls. }
\end{abstract}

\maketitle

Crystals mainly deform through the motion of dislocations~\cite{Asaro:2006lh}, extended line-like defects in the crystal lattice. This sole fact can provide key explanations for the 
magnitude and character of uniaxial and shear strength, as well as the plastic crystalline behavior. 
However, the complex spatio-temporal dynamics of dislocations is elusive:  It has long been known that many aspects of the stress-strain response of deformed 
metals are associated to collective and emergent
dislocation patterning that cannot be predicted by single-dislocation features, e.g., dense dipolar wall (DDW) formation and corresponding dislocation cell walls (see for 
example, 
Refs.~\cite{Chen:2010kl,Mughrabi:1986tg,Schwink:1992hc,Hahner:1998ij,Hughes:1997bs,Hughes:1998fv,
Laufer:1966dz}). Also, failure due to mechanical fatigue is preceded by the 
formation of complex dislocation patterns that have been labeled as ``vein structures", typically 
observed after multiple thousands of fatigue 
cycles~\cite{Suresh:1998fu,Mughbrabi:1979kl,Walgraef:1985qa,Kuhlmann-Wilsdorf:1980tx,Tabata:1983lk}. 
Dislocation patterning is, therefore, not only interesting in relation to analogous 
phenomena in statistical mechanics~\cite{Goldenfeld:1992aa}, but also it correlates with the vital 
technological interest of characterizing and predicting the lifetime of mechanical 
components~\cite{Suresh:1998fu}.

\begin{figure}[t!]
\begin{picture}(0,0)
\put(-5,80){\sffamily{(a)}}
\end{picture}
\includegraphics[angle=0, width=3.5cm]{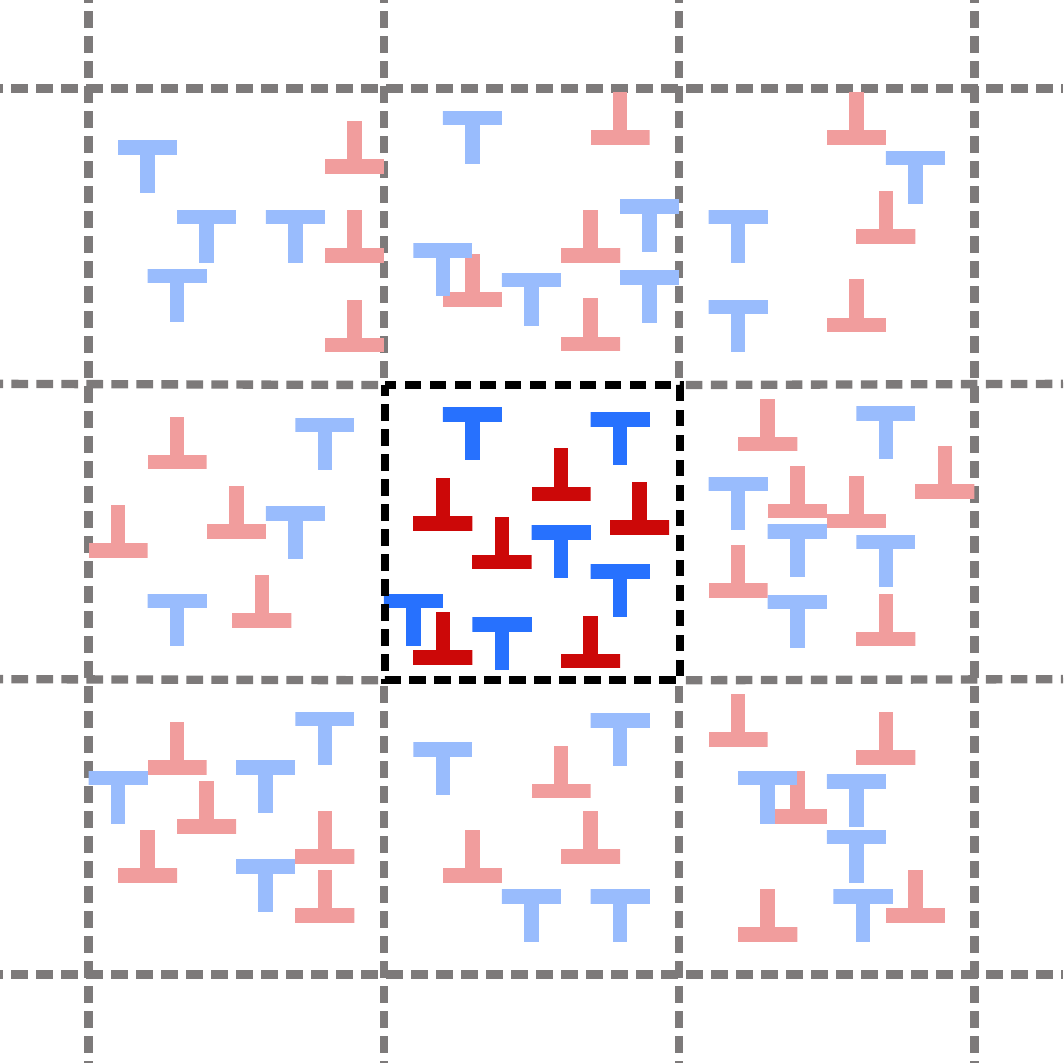}
\hspace*{0.5cm}
\begin{picture}(0,0)
\put(-5,80){\sffamily{(b)}}
\end{picture}
\includegraphics[angle=0, width=3.5cm]{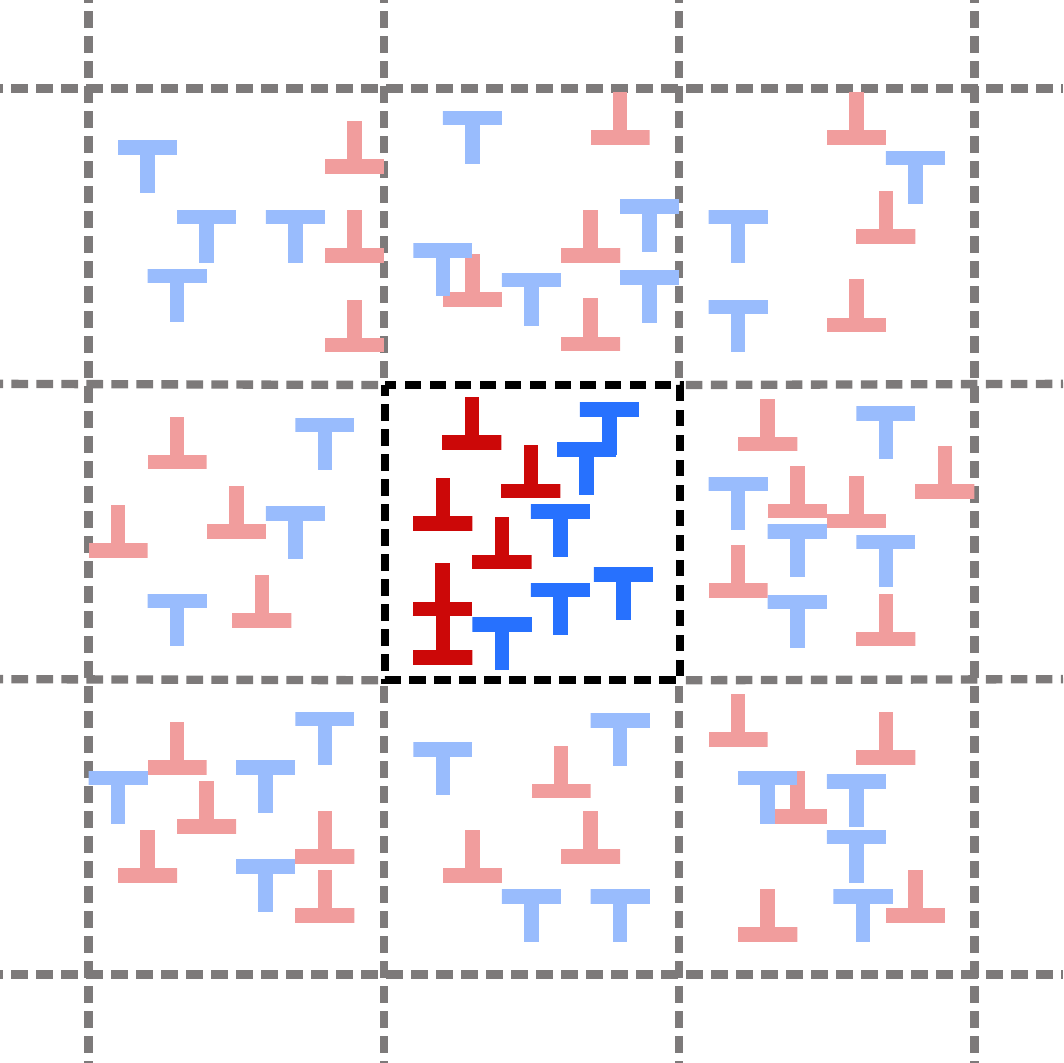}
\caption{A gradient in the geometrically necessary dislocation density is one form of spatial correlations that may affect local strength (through the back-stress $\tau_b$) and lead to different yielding thresholds for panels (a) and (b). Configuration of panel (b) is stronger (weaker) if $D<0$ ($D>0$). The dashed lines denote spatial discretization which is necessary to define continuous densities.
\vspace*{-5mm}
\label{fig:sketch}}
\end{figure}

Although during the past five decades several attempts have been proposed to 
model pattern 
formation~\cite{Holt1970_JAP,Aifantis1985_JAP,Pontes2006_IJP,Salazar1995_AM,Hansen1986_MSE,
kratochvil2008} most of them are based on purely phenomenological arguments, so they are not derived 
from the properties of individual dislocations. Very recently, however, models based on statistical 
physics considerations have been derived establishing a direct link between a micro and mesoscale 
descriptions of the collective motion of dislocations \cite{Groma:2003pi, groma_debye_2006, groma_variational_2010, groma_scale-free_2015, groma_dislocation_2016, Valdenaire2016_PRB}.
Yet, neither the dynamical or energetic origin of DDWs (or that of the vein structures or other specific 
dislocation patterns) was clarified, nor their relation to 
\emph{back-stress} terms (induced by dislocation correlations schematically shown in Fig.~\ref{fig:sketch}) appearing in continuum dislocation dynamics. These latter dislocation force 
components are typically associated with the Bauschinger effect (the asymmetry of deformation upon reversed loading) and represent key ingredients of 
kinematic hardening theories~\cite{Prager:1955ye,Asaro:2006lh,Prager:1956qo}. In dislocation 
dynamics and strain-gradient plasticity theories~\cite{Fleck:1994aa}, such terms involve non-linear 
derivatives of the local dislocation density, however their precise form has been 
elusive~\cite{Groma:2003pi,Zaiser:2007le,dogge_extended_2015}.

In this paper, we develop a stochastic continuum model  based on the deterministic 
approach proposed by Groma et.~al.~\cite{groma_dislocation_2016} that is able to display the spontaneous formation of 
dislocation walls  through a dynamical transition~\cite{Goldenfeld:1992aa}, and can be used to 
establish basic constitutive rules for continuum dislocation plasticity theories. The key ingredients 
are: a particular form of dislocation backstress, a ''diffusion`` like stress term proportional to the gradient of the statistically stored dislocation density, and the flow stress being proportional to the square root of the statistically stored dislocation density (Taylor relation). Each terms  arise  from a consistent coarse-graining 
procedure and contain  dimensionless prefactors (for details see below). We show that realistic DDW formation occurs 
only when the back and diffusion stresses are obtained from a convex functional, and in this case our continuum model becomes consistent with lower scale discrete dislocation 
dynamics (DDD) simulations~\cite{zhou_dynamic_2015}. Our results, therefore, shed new light on the 
origin of the backstress as well as its role in dislocation patterning of bulk single crystals, and 
provide a successful multi-scale description of the dynamics in single-slip edge dislocation 
systems.

\begin{table}[t]
\caption{\label{tab:units}Summary of the units of dimensionless quantities.}
\begin{ruledtabular}
\begin{tabular}{lc}
Quantity & Unit \\
\hline
Distance ($x$) & $\rho_0^{-1/2}$  \\
Stress ($\tau$) & $Gb\rho_0^{1/2}$ \\
Plastic strain ($\gamma$) & $b\rho_0^{1/2}$ \\
Time ($t$) & $MGb^2\rho_0$ \\
Dislocation density ($\rho, \kappa$) & $\rho_0$ \\
\end{tabular}
\end{ruledtabular}
\vspace*{-0.4cm}
\end{table}

The emergence of dislocation patterns has been investigated numerically using multiple approaches, including two-dimensional(2D)~\cite{zhou_random_2014, zhou_dynamic_2015, Ispanovity:2014ff, Ispanovity:2010mi, szabo_plastic_2015, laurson_dynamical_2010, rosti_fluctuations_2010, papanikolaou_obstacles_2017, gomez-garcia_dislocation_2006}, three-dimensional (3D) DDD~\cite{papanikolaou_avalanches_2017, devincre_dislocation_2008}, as well as continuum dislocation dynamics (CDD)~\cite{el-azab_statistical_2000, xia_computational_2015, chen_bending_2010, chen_scaling_2013}. Realistic 3D-DDD have been too expensive and remain below $1\%$ strain in bulk conditions, while CDD has not yet captured local entanglement and backstress interactions that are expected to play critical role in patterning~\cite{hochrainer_continuum_2014, hochrainer_thermodynamically_2016}. In contrast, 2D-DDD methods not only are numerically tractable at large strains,  but also a rigorous coarse graining procedure has been developed for the special case of edge dislocations in single slip~\cite{Groma:2003pi, groma_variational_2010, groma_dislocation_2016}. In this paper, we will investigate the continuum description of patterns in this case.

We consider a configuration of straight parallel edge dislocations, lying along the $z$ axis with their Burgers vectors pointing in the $x$ direction. We assume to track the motion of dislocations on the $z=0$ plane. To emulate an infinite crystalline medium, periodic boundary conditions (PBC) are applied at the borders of the square shaped simulation area of size $L \times L$. The Burgers vector can be written as $\bm b_i = s_i \bm b$, where $\bm b = (b, 0)$, $s_i = \pm 1$, and $1 \le i \le N$, with $N$ being the total number of dislocations. Throughout of this paper dimensionless units summarized in Table \ref{tab:units} are used where $\rho_0 = N/L^2$ is the average total dislocation density and $G=\mu/[2\pi(1-\nu)]$ is an elastic constant (where $\mu$ and $\nu$ is the shear modulus and Poisson's number, respectively). To mimic easy glide a linear relationship is assumed between the projection of the Peach-Koehler force to the glide plane ( denoted by $F$ hereafter) and the dislocation velocity in the glide direction $v$ as $v = M F$, where $M$ is the dislocation mobility~\cite{amodeo_dislocation_1990}. For further details  on our 2D-DDD simulations see the Supplemental Material.

\begin{figure}[!t]
\begin{center}
\vspace*{1.0cm}
\begin{picture}(0,0)
%\put(-3,95){\sffamily{(a)}}
\put(-2,29){\rotatebox{90}{\sffamily{$\gamma=0$}}}
\put(10,95){\parbox{1in}{\sffamily{Discrete\\ configuration}}}
\put(85,95){\parbox{1in}{\sffamily{Same sign correlation function ($d_{++}$)}}}
\put(154,95){\parbox{1in}{\sffamily{Opposite sign correlation function ($d_{+-}$)}}}
\end{picture}
\hspace*{2mm}
\includegraphics[angle=0, height=3.1cm, trim=25mm 0 22mm 0]{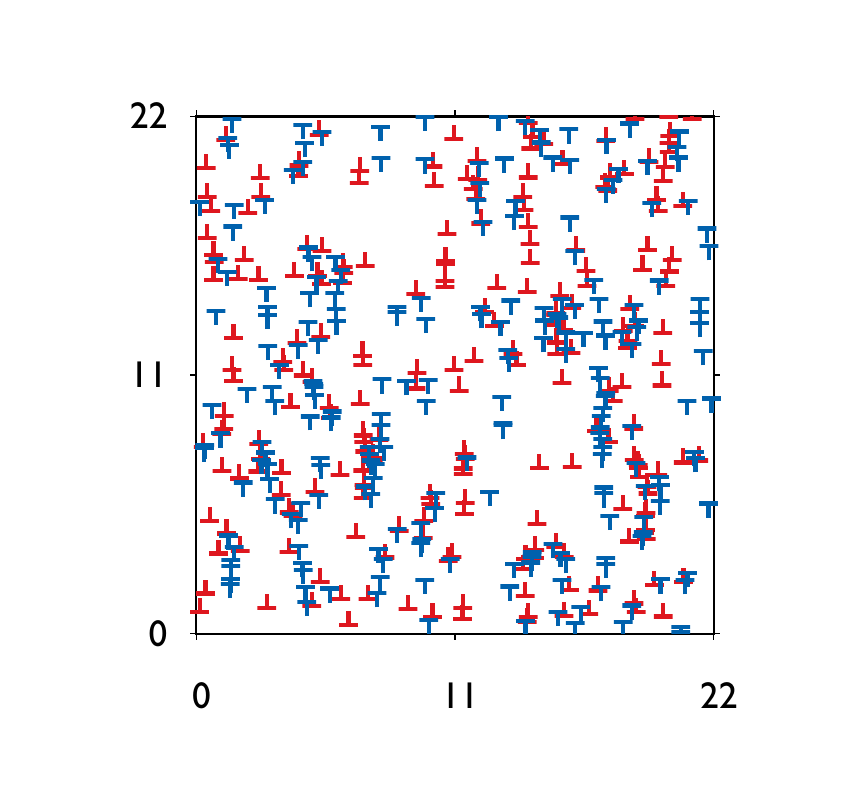}
\includegraphics[angle=0, height=3.1cm, trim=12mm 0 8mm 0]{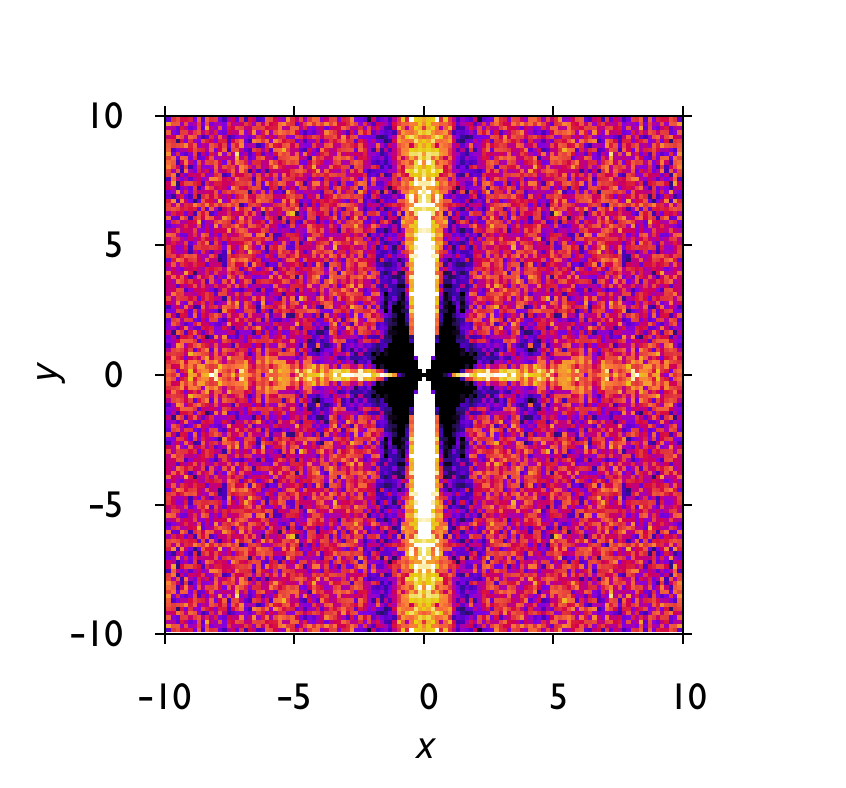}
\includegraphics[angle=0, height=3.1cm, trim=35mm 0 8mm 0]{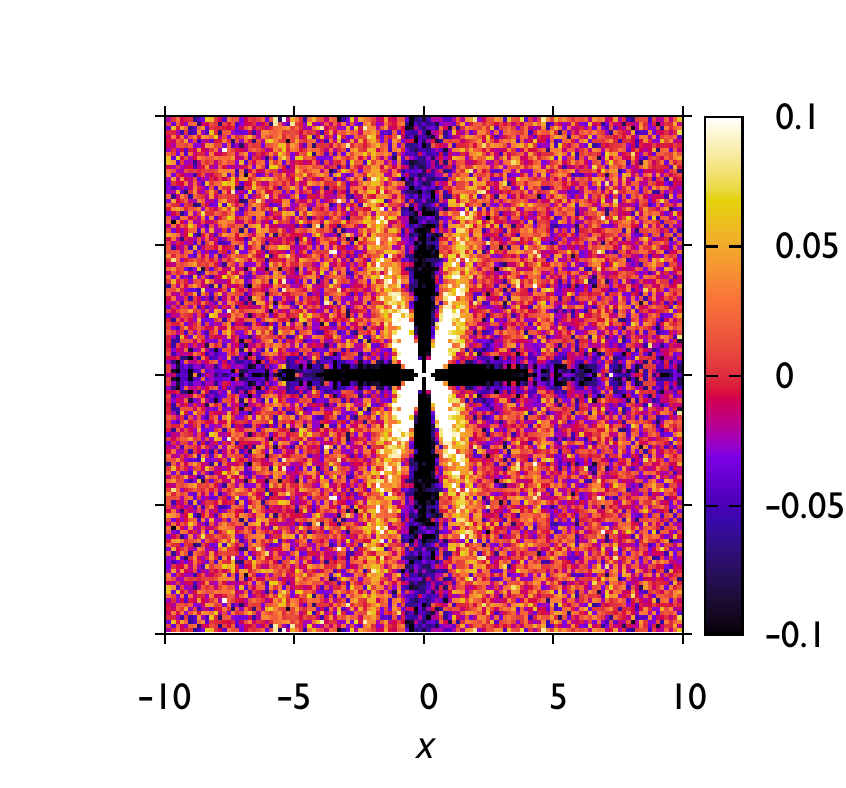}
\vspace*{-2mm}

\vspace*{-2mm}
\begin{picture}(0,0)
\put(-2,28){\rotatebox{90}{\sffamily{$\gamma=0.4$}}}
\end{picture}
\hspace*{2mm}
\includegraphics[angle=0, height=3.1cm, trim=25mm 0 22mm 0]{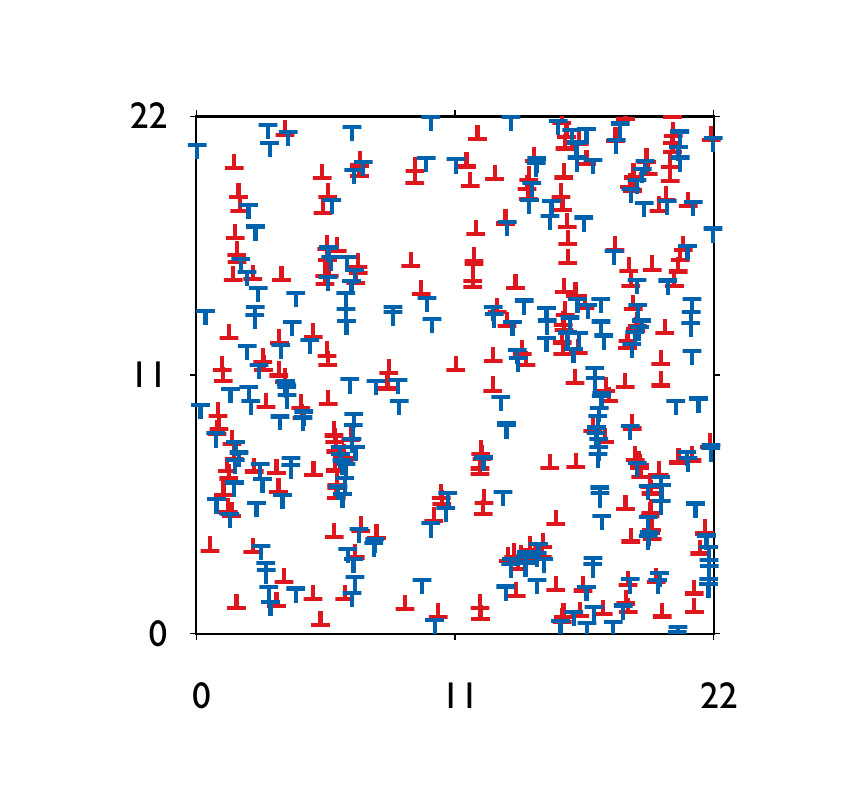}
\includegraphics[angle=0, height=3.1cm, trim=12mm 0 8mm 0]{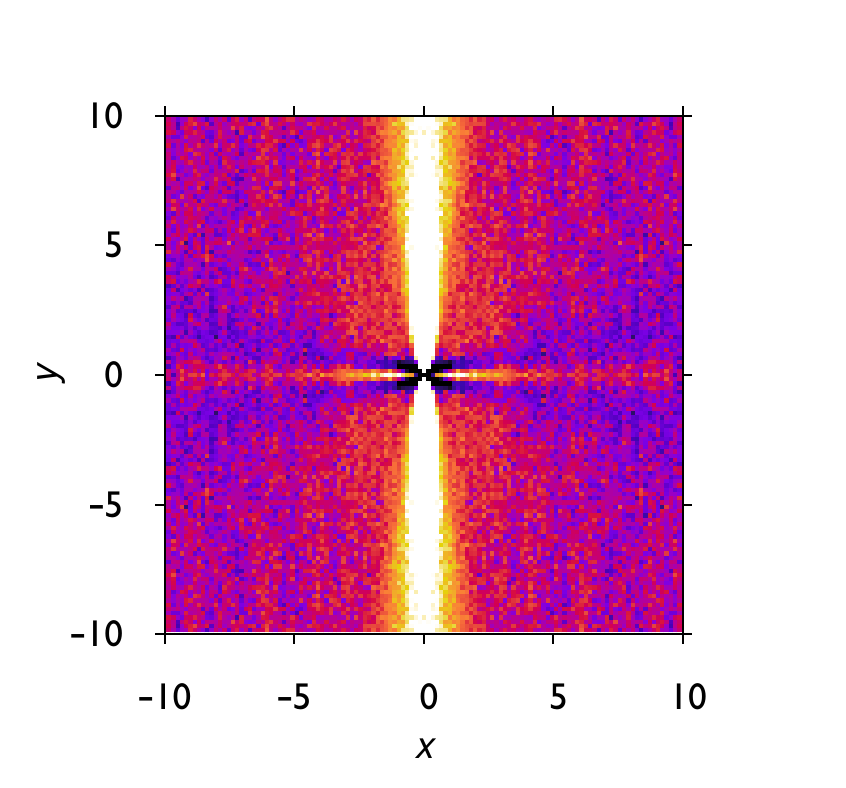}
\includegraphics[angle=0, height=3.1cm, trim=35mm 0 8mm 0]{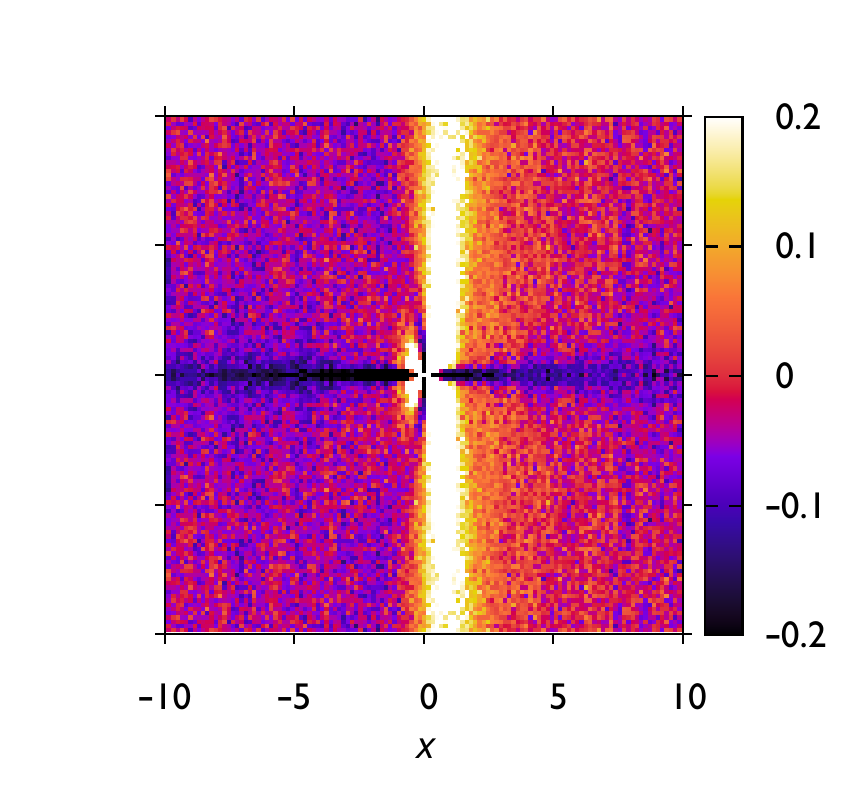}
\vspace*{-2mm}

\vspace*{-2mm}
\begin{picture}(0,0)
\put(-2,28){\rotatebox{90}{\sffamily{$\gamma=1$}}}
\end{picture}
\hspace*{2mm}
\includegraphics[angle=0, height=3.1cm, trim=25mm 0 22mm 0]{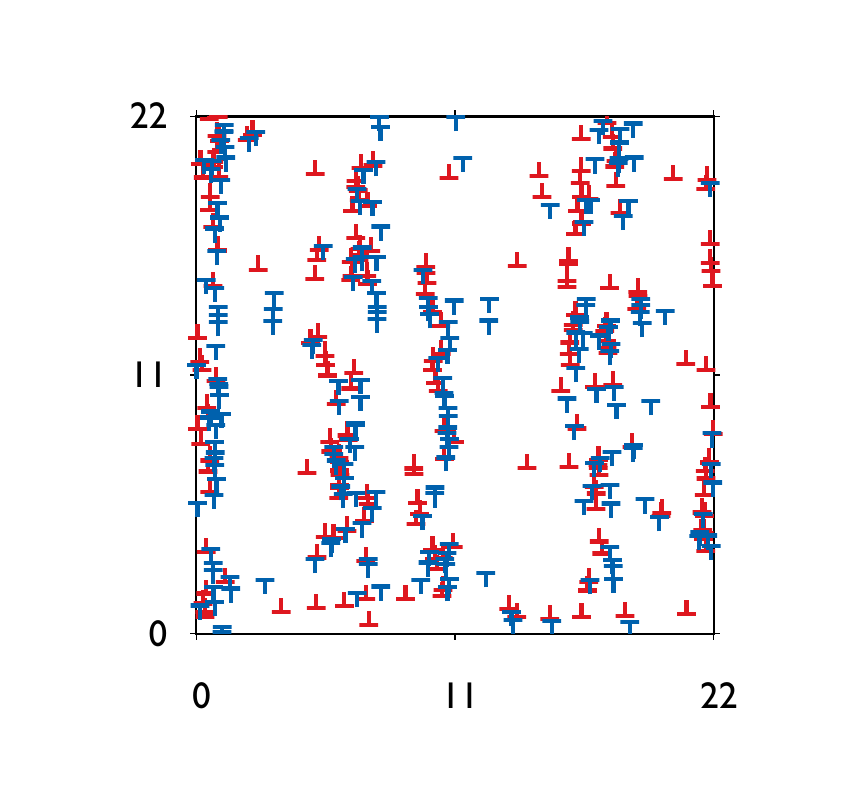}
\includegraphics[angle=0, height=3.1cm, trim=12mm 0 8mm 0]{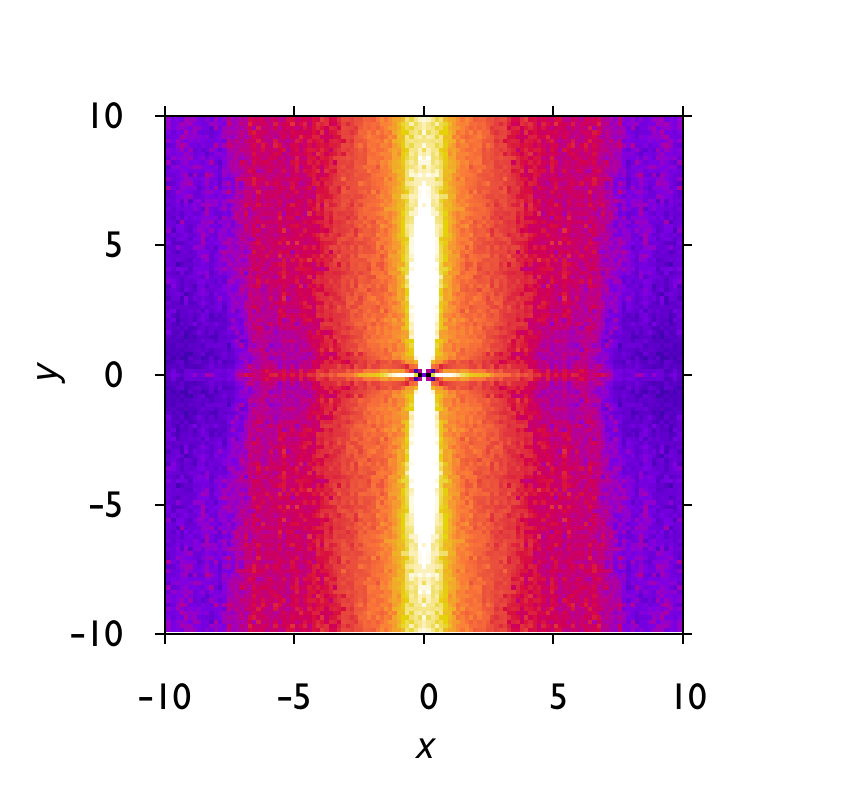}
\includegraphics[angle=0, height=3.1cm, trim=35mm 0 8mm 0]{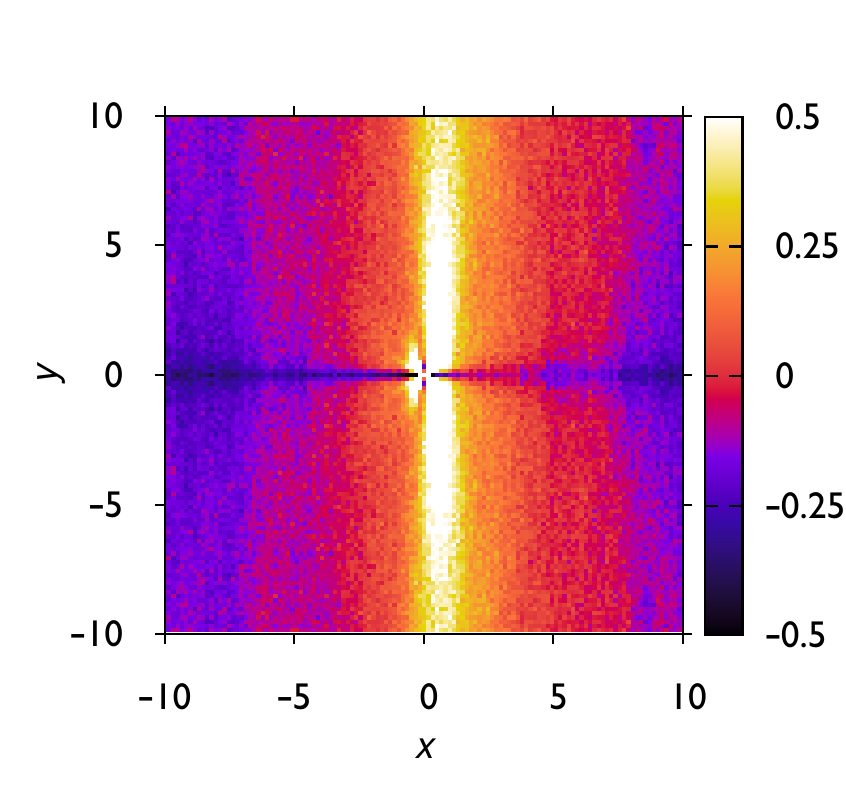}
\vspace*{-2mm}

\vspace*{-2mm}
\begin{picture}(0,0)
\put(-2,28){\rotatebox{90}{\sffamily{$\gamma=32$}}}
\end{picture}
\hspace*{2mm}
\includegraphics[angle=0, height=3.1cm, trim=25mm 0 22mm 0]{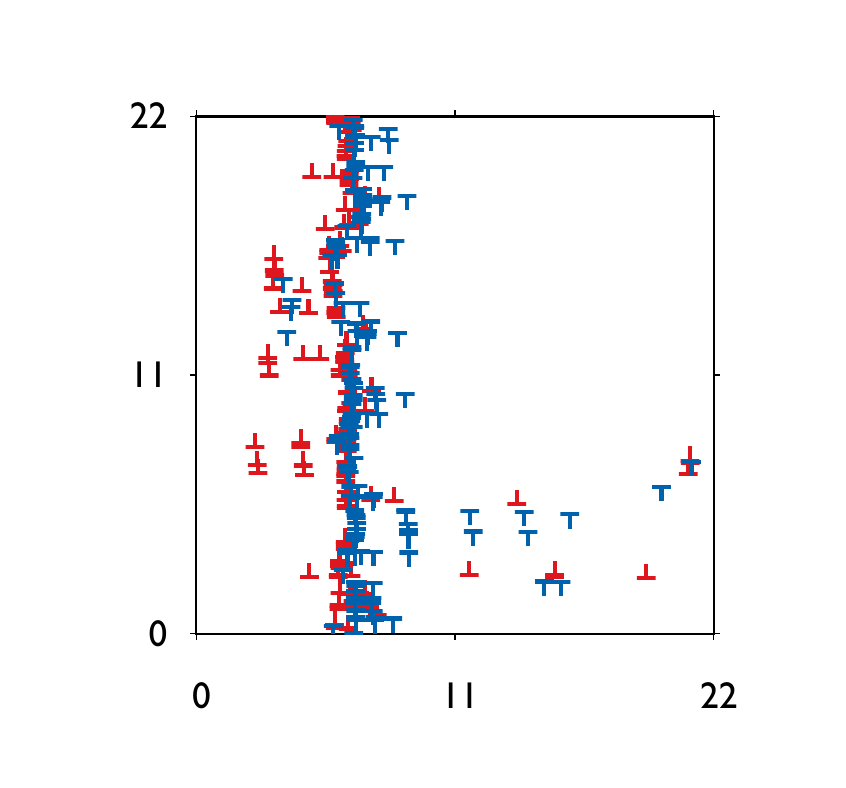}
\includegraphics[angle=0, height=3.1cm, trim=12mm 0 8mm 0]{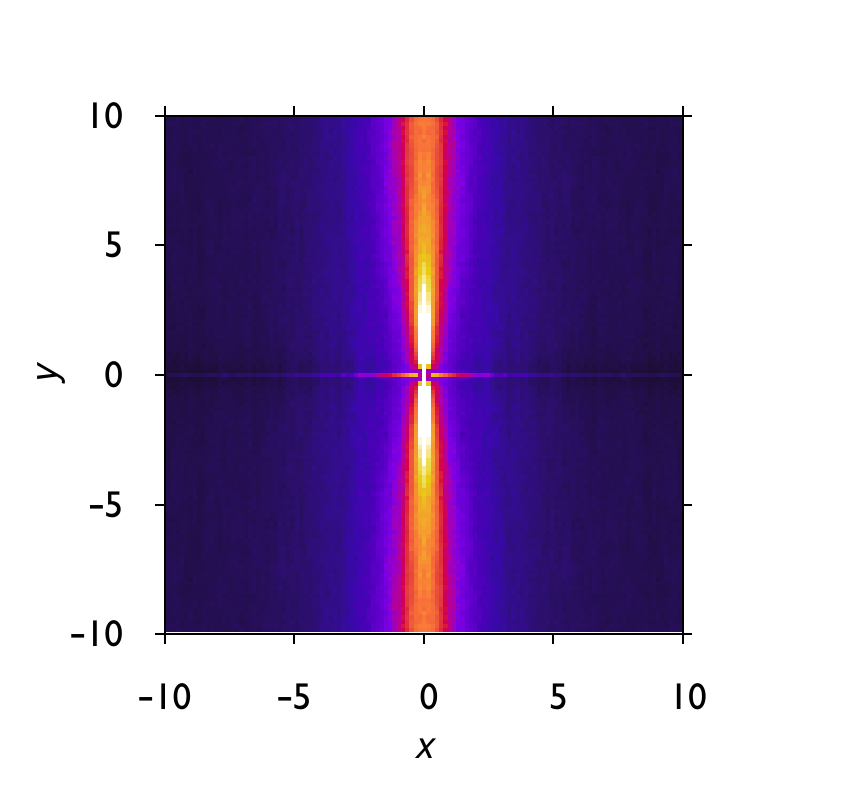}
\includegraphics[angle=0, height=3.1cm, trim=35mm 0 8mm 0]{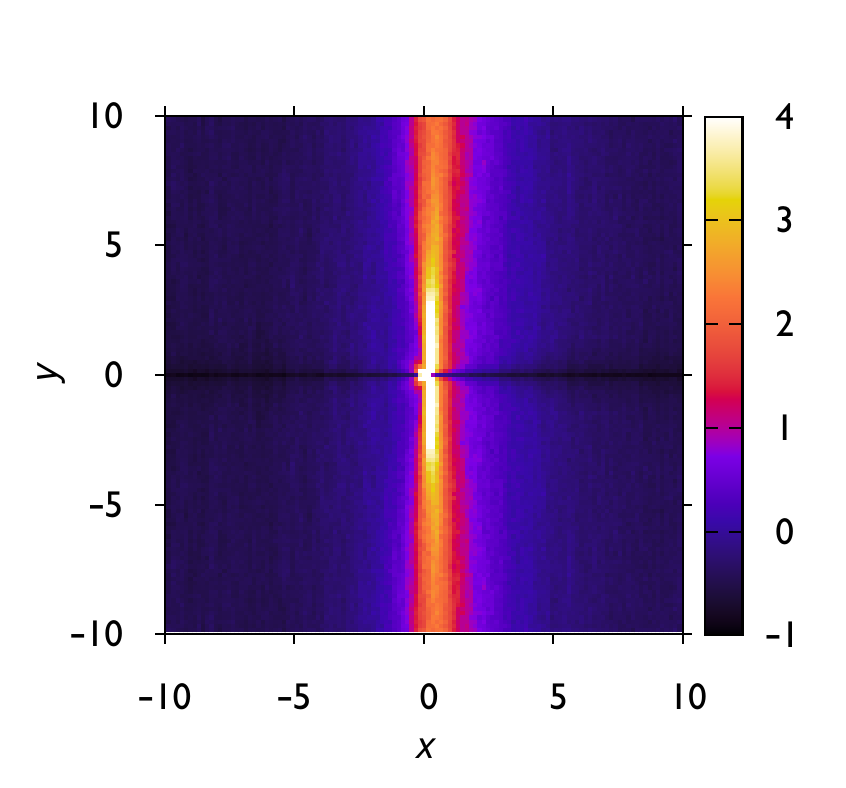}
\vspace*{-2mm}

%
%\hspace*{-10mm}
%\begin{picture}(0,0)
%\put(-12,100){\sffamily{(b)}}
%\end{picture}
%\includegraphics[angle=0, height=4cm, trim=4mm 0 8mm 0]{Figures/stress_strain/stress_strain_comparison.pdf}
\end{center}
\vspace*{-4mm}
\caption{Dislocation pattern evolution in DDD simulations. Left column: Dislocation configurations obtained at different strain $\gamma$ values. Middle and right columns: Same sign and opposite sign spatial correlation functions ($d_{++}$ and $d_{+-}$, respectively) of discrete dislocations.
\vspace*{-0.7cm}
\label{fig:pattern_evol_ddd}}
\end{figure}

The typical evolution of a dislocation configuration can be seen in the left column of Fig.~\ref{fig:pattern_evol_ddd}. At zero applied shear stress $\tau_{\rm ext}$ no clear pattern can be observed even though there are specific local (low energy) configurations: Opposite sign dislocations organize into short dipoles whereas those of identical sign form short vertical walls. As $\tau_{\rm ext}$ increases, dislocation patterns become increasingly heterogeneous with predominant long dense vertical walls~\cite{zhou_dynamic_2015}. These DDWs are induced by the  external shear stress and represent the most stable configuration that can be formed in this 2D system.
%The stress-strain curve corresponding to this process is seen in Fig.~\ref{fig:pattern_evol_ddd}(b) (see also~\cite{zhou_dynamic_2015, szabo_plastic_2015}).
Recently, it has been shown that orientation of the slip system with respect to the simulation box strongly influences the correlation properties of the dislocation network at large strains \cite{kapetanou_stress_2015}. This type of boundary condition sensitivity is common to patterning instabilities in condensed matter systems with long-range interactions~\cite{luijten_universality_2002}. In all such systems, the local interaction that causes the instability is believed to be independent of the particular boundary condition to be investigated. Thus, in the present system, the emergent local order is not expected to be affected by boundary conditions at small strains.

To characterize the pattern evolution in a quantitative manner the two-point spatial correlation functions are computed at different strain levels. These functions are defined as
\begin{equation}
	d_{++}(\bm r) = \left\langle \frac{\rho_{++}(\bm r', \bm r' + \bm r)}{\rho_0^2}-1 \right\rangle_{\bm r'},
\end{equation}
where $\rho_{++}(\bm r', \bm r)$ denotes the two-point density of a $+$ sign dislocation at $\bm r'$ and a $+$ sign dislocation at $\bm r' + \bm r$ \cite{zaiser_statistical_2001}. In the definition the averaging with respect to $\bm r'$ is introduced since the system due to the PBC is homogeneous, so, $\rho_{++}(\bm r_1, \bm r_2)$ can only depend on the relative coordinate $\bm r_2 - \bm r_1$. The correlation function for the opposite sign dislocations $d_{+-}$ is defined accordingly. The numerically obtained correlation functions and their evolution with increasing strain can be seen in the center and right column of Fig.~\ref{fig:pattern_evol_ddd}. These functions were averaged over 200 individual realizations, so, they represent the average patterning behavior of the DDD system. At zero strain the microstructure is dominated by vertical walls of same sign dislocations and short dipoles of opposite sign dislocations having a $45^\circ$ angle with the horizontal glide plane. The $d_{++}$ function has a slow $|y|^{-1.5}$ type power-law decay along the $y$ axis, signalling a broad distribution of the wall lengths \cite{groma_debye_2006}. With increasing strain
\begin{itemize}
\item[(i)] the walls get longer and denser (see the change in the magnitude of the correlation function $d_{++}$) and
\item[(ii)] the function $d_{+-}$ becomes asymmetric due to the polarization of the microstructure, that is, the $-$ sign dislocations tend to be positioned to the right with respect to the $+$ sign dislocations. In addition, with accumulating strain the $d_{+-}$ function extends in the vertical direction similarly to the $d_{++}$ signalling the build-up of the DDWs seen in the left column of Fig.~\ref{fig:pattern_evol_ddd}.
\end{itemize} 

In order to identify the precise continuum form of the DDW instability in the 2D-DDD simulations, we consider the theory that has been directly derived from the equations of motion (see SI) using a rigorous coarse graining procedure \cite{groma_dislocation_2016}, and is based on the continuous density fields $\rho_\pm(\bm r, t)$ of dislocations with identical ($+$ or $-$) sign, and the corresponding total dislocation density $\rho = \rho_+ + \rho_-$ and geometrically necessary dislocation (GND) density $\kappa = \rho_+ - \rho_-$. The recently revisited form of the evolution equations are as follows:
\begin{gather}
 \partial_t \rho_+=  -\partial_x 
\left\{\rho_+\left[ \tau_\text{ext} + \tau_\text{sc} + \tau_b-2\frac{\rho_-}{\rho}\tau_f+\tau_d 
\right]\right\}, \label{eq:rhod2plus}\\
 \partial_t \rho_-= +\partial_x 
\left\{\rho_-\left[ \tau_\text{ext} + \tau_\text{sc} +\tau_b-2\frac{\rho_+}{\rho}\tau_f-\tau_d 
\right]\right\}, \label{eq:rhod2minus}
\end{gather}
where
\begin{align}
\tau_\text{sc}(\bm r, t) &= \int \tau_{\rm ind}(\bm{r}-\bm{r}')\kappa(\bm{r}', t){\rm d}^2r'
\end{align}
is the long-range (or ``self-consistent'') stress field of GNDs which together with the external field $\tau_\text{ext}$ represents the experimentally measurable average shear stress in a small volume around $\bm r$. This is complemented by the  gradient stress components
\begin{equation}
\tau_b(\bm r, t) = -\frac{D}{\rho}\partial_x\kappa(\bm r, t) \, \text{ and } \,
\tau_d(\bm r, t)= -\frac{A}{\rho}\partial_x\rho(\bm r, t).  \label{eq:gradient_stress}
\end{equation}
and friction stress $\tau_f$ that is as big as necessary to prevent dislocation flow and it cannot be larger than the flow stress $\tau_\text{flow} = \alpha \rho^{1/2}$. In the equations above, $\alpha$, $D$ and $A$ are dimensionless constants.

The origin of the friction stress $\tau_f$ and local gradient terms $\tau_b$ (back-stress) and $\tau_d$ (diffusion stress) is clear from the formal derivation of the theory \cite{groma_dislocation_2016}: They stem from the fact that dislocations are not positioned randomly but are spatially correlated, a fact that has been already postulated by Wilkens based on energetic considerations \cite{wilkens_mittlere_1969}, also demonstrated in numerical simulations \cite{zaiser_statistical_2001} and in Fig.~\ref{fig:pattern_evol_ddd}. Dislocation patterns themselves are also a manifestation of these correlations (see, for example, Fig.~1 for a schematic). As of the physical meaning of these terms, friction stress, and so the flow stress, is the result of the small-scale correlated substructures (most importantly, dislocation dipoles in the 2D system being studied) that may be stable against external load.  Indeed, in Eqs.~(\ref{eq:rhod2plus},\ref{eq:rhod2minus}) $\tau_f$ is multiplied by $\rho_\pm$ expressing that dislocations can only be withheld by dislocations of opposite sign. Interpretation of gradient terms are more subtle: In the flowing regime they can be envisaged as a correction to the flow stress. In particular, due to the back-stress term local strength may depend on the gradient of the GND density as depicted in the sketch of Figure~\ref{fig:sketch}. According to the sign of parameter $D$, the strength of the local volume in Fig.~\ref{fig:sketch}(b) may be larger (for $D<0$) or smaller (for $D>0$) than that of Fig.~\ref{fig:sketch}(a). Similar explanation can be given for the diffusion stress $\tau_d$. 

\begin{figure}[tbh]
\hspace*{-0.5cm}
%\begin{picture}(0,0)
%\put(25,95){\sffamily{(a)}}
%\end{picture}
\includegraphics[width=0.25\textwidth]{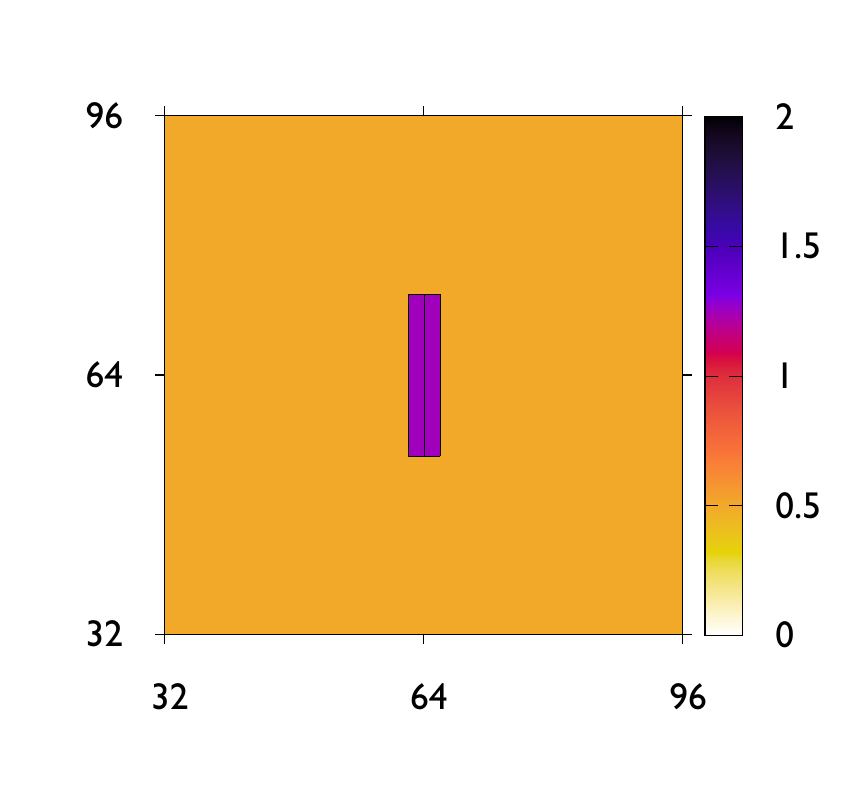}
\begin{picture}(0,0)
\put(-105,90){\colorbox{white!20}{\sffamily{(a)}}}
\end{picture}
\hspace*{-0.5cm}
\includegraphics[width=0.25\textwidth]{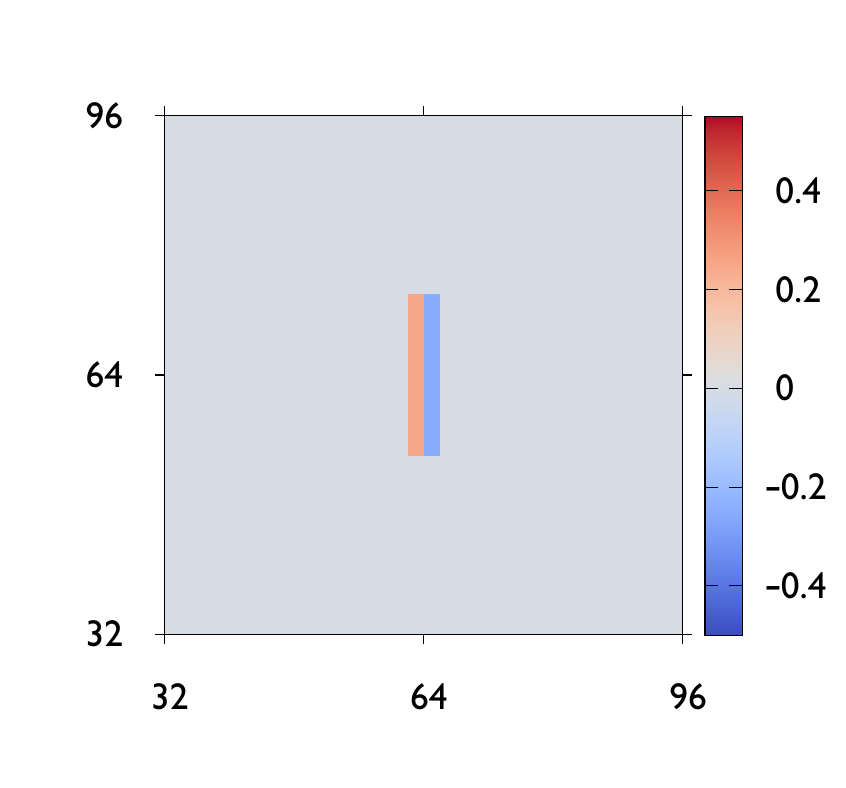}
\begin{picture}(0,0)
\put(-105,90){\colorbox{white!20}{\sffamily{(b)}}}
\end{picture}
\vspace*{-0.5cm}

\hspace*{-0.5cm}
\includegraphics[width=0.25\textwidth]{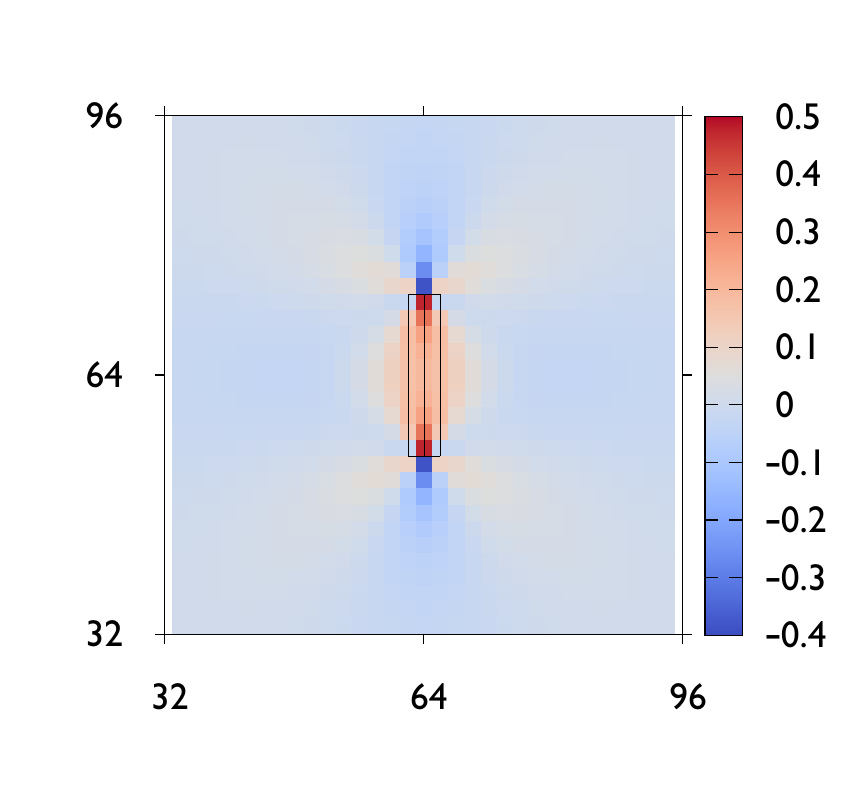}
\begin{picture}(0,0)
\put(-105,90){\colorbox{white!20}{\sffamily{(c)}}}
\end{picture}
\hspace*{-0.5cm}
\includegraphics[width=0.25\textwidth]{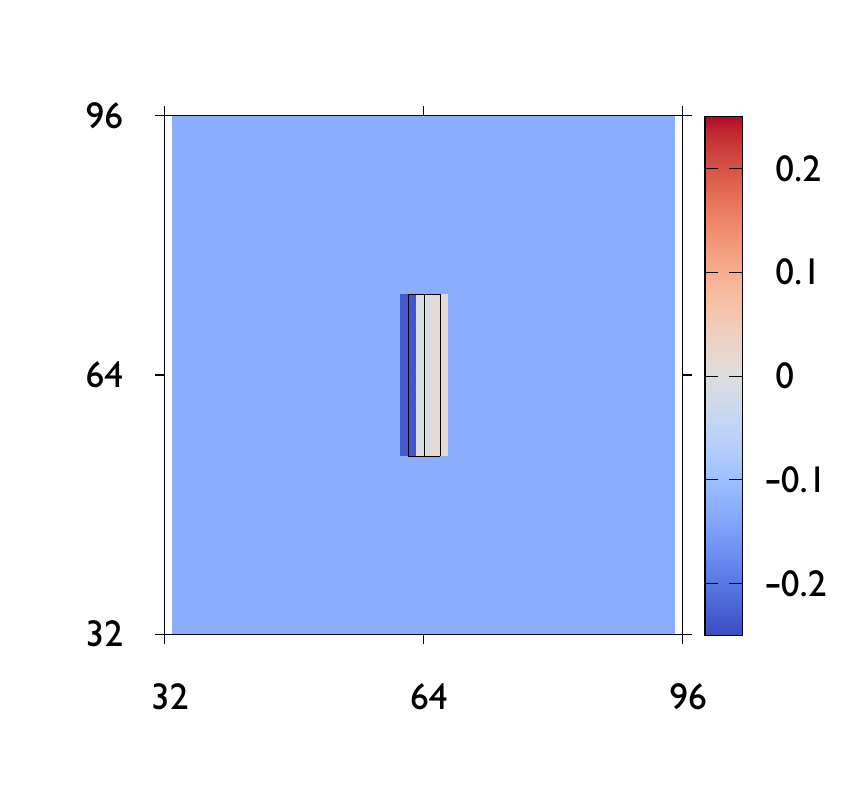}
\begin{picture}(0,0)
\put(-105,90){\colorbox{white!20}{\sffamily{(d)}}}
\end{picture}
\caption{
Stress fields in a model of a dipolar wall in the SCDD. (a,b):
Total and GND densities of a dipolar wall placed in a constant
dislocation density background. (c) Self-consistent field
$\tau_\text{sc}$ of the dipolar wall. (d) Gradient stresses ($\tau_b +
\tau_d$) acting on a positive sign ($s=1$) dislocation. Notice the
development of a back-stress on the left side of the dipolar wall. The model parameters used are the same as in Fig.~\ref{fig:pattern_evol_continuum}.}
\label{fig:bck}
\end{figure}

The importance of our formulation becomes transparent when one aims at building a multiscale connection and understanding of traditional plasticity theories of constitutive formulations of backstress fields to discrete dislocation formulations~\cite{Asaro:2006lh}. In traditional formulations, the most common approach involves the modeling of backstress $\chi$ in a particular slip system, as a correction to the shear stress $\tau$, that displays a typical evolution equation with direct hardening and dynamic recovery coefficients respectively~(see for example Ref.~\cite{Morrissey:2001fk}):
\bea
\dot\chi = c\dot\gamma - d \chi |\dot\gamma|
\label{eq:cd}
\eea
where $\gamma,\dot\gamma$ are the shear strain and strain rates in the slip system. This equation has a natural, intuitive and experimentally verifiable explanation in terms of polycrystalline, grain-boundary dominated plasticity, since it is transparent that dislocations that ``cross" grain-boundaries should face an opposite-signed, discontinuous resistance when loaded in the opposite direction.
However, in dislocation ensembles, such explanations were elusive and vague, especially since continuum dislocation density theories desire continuous, analytic terms if not required otherwise. Nevertheless, in the single-slip theory we have developed, it is transparent how the $d$-term in Eq.~(\ref{eq:cd}) emerges, without the particular need of non-analytic mathematical terms. As shown in Fig.~\ref{fig:bck}, the response of a continuum dipolar wall in our theory contains all the ingredients that were constitutively assumed in traditional multiscale formulations: If a positive-signed dislocation moves across the wall, it would have an opposite-signed resistance if moved in opposite directions.

\begin{figure*}[!ht]
\begin{center}
\vspace*{1.0cm}
\begin{picture}(0,0)
%\put(-3,95){\sffamily{(a)}}
\put(-10,29){\rotatebox{90}{\sffamily{$\gamma=0$}}}
\put(8,95){\parbox{1in}{\sffamily{Total density\\($\rho$)}}}
\put(90,95){\parbox{1in}{\sffamily{GND density\\($\kappa$)}}}
\put(185,95){\parbox{1in}{\sffamily{Same sign correlation function ($d_{++}$)}}}
\put(258,95){\parbox{1in}{\sffamily{Opposite sign correlation function ($d_{+-}$)}}}
\end{picture}
\hspace*{2mm}
\includegraphics[angle=0, height=3.1cm, trim=25mm 0 22mm 0]{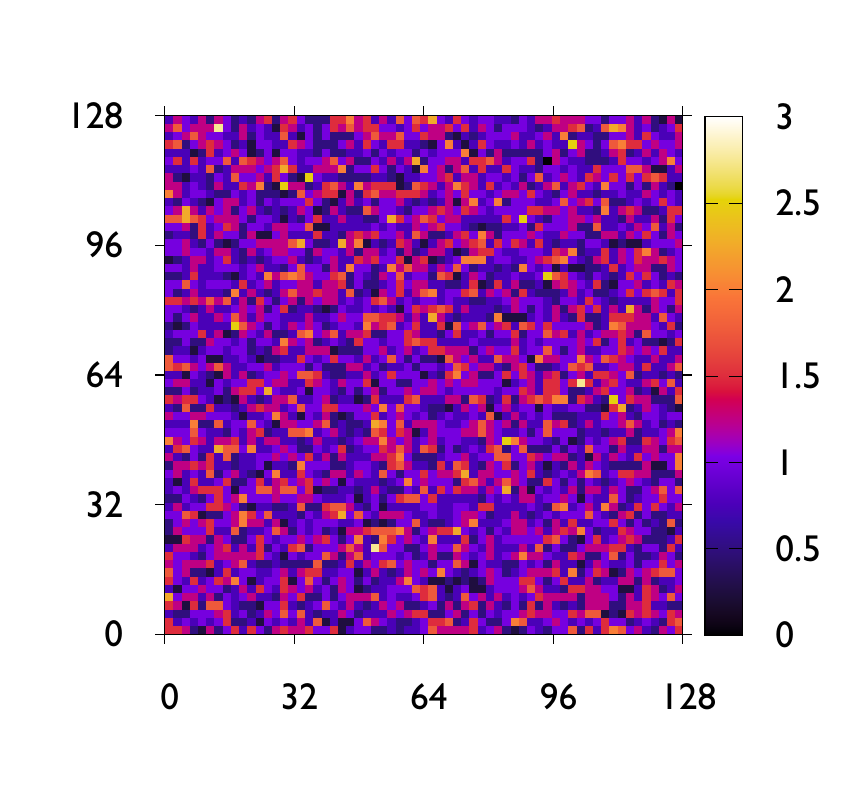}
\includegraphics[angle=0, height=3.1cm, trim=5mm 0 0 0]{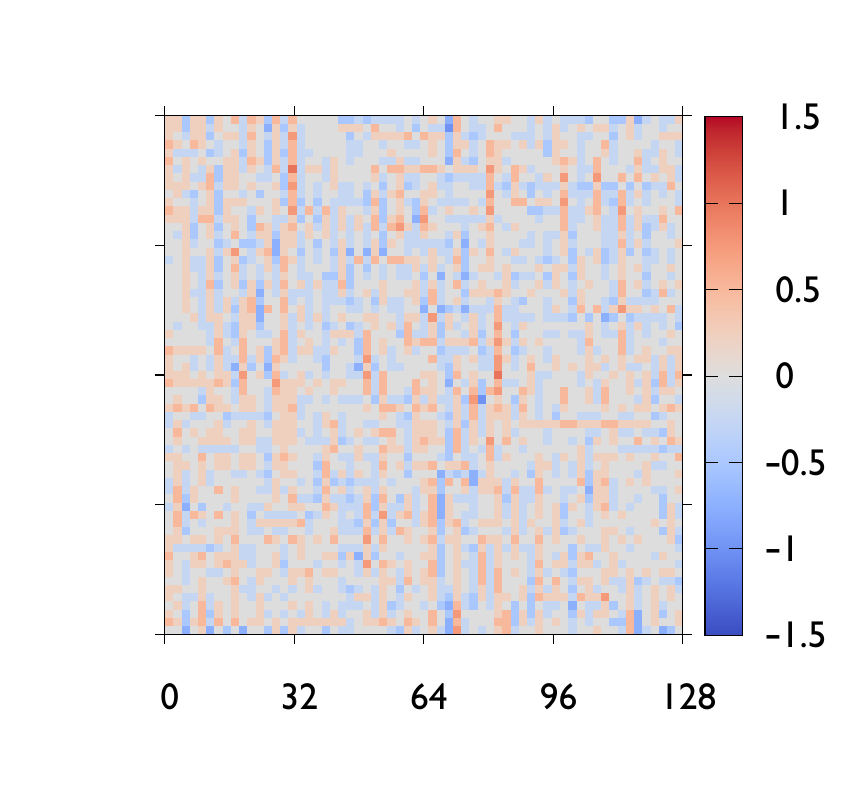}
\includegraphics[angle=0, height=3.1cm, trim=0mm 0 8mm 0]{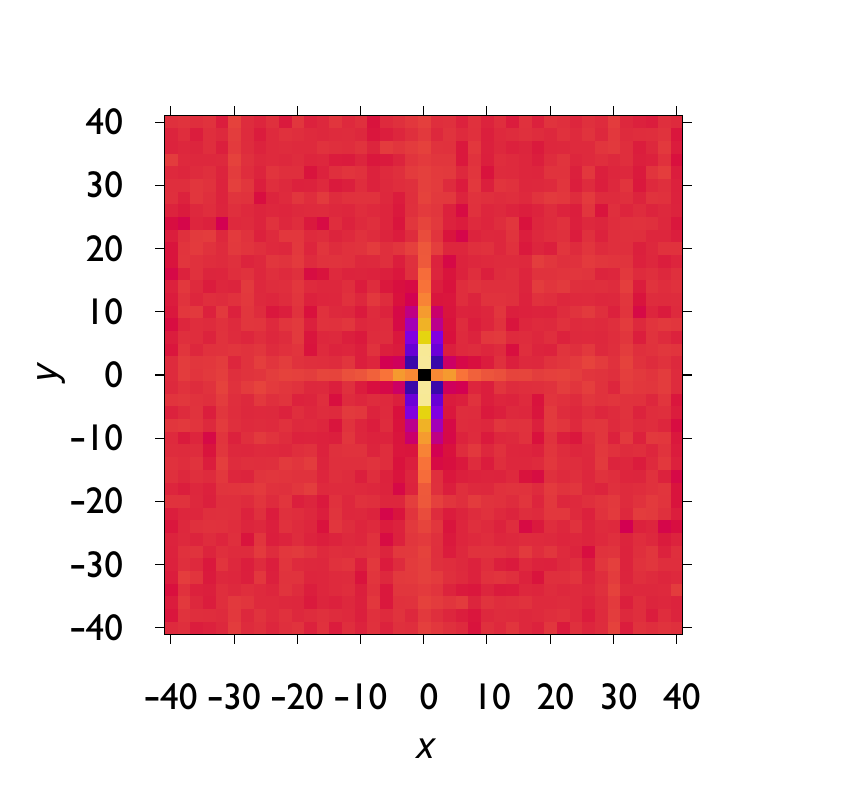}
\includegraphics[angle=0, height=3.1cm, trim=35mm 0 8mm 0]{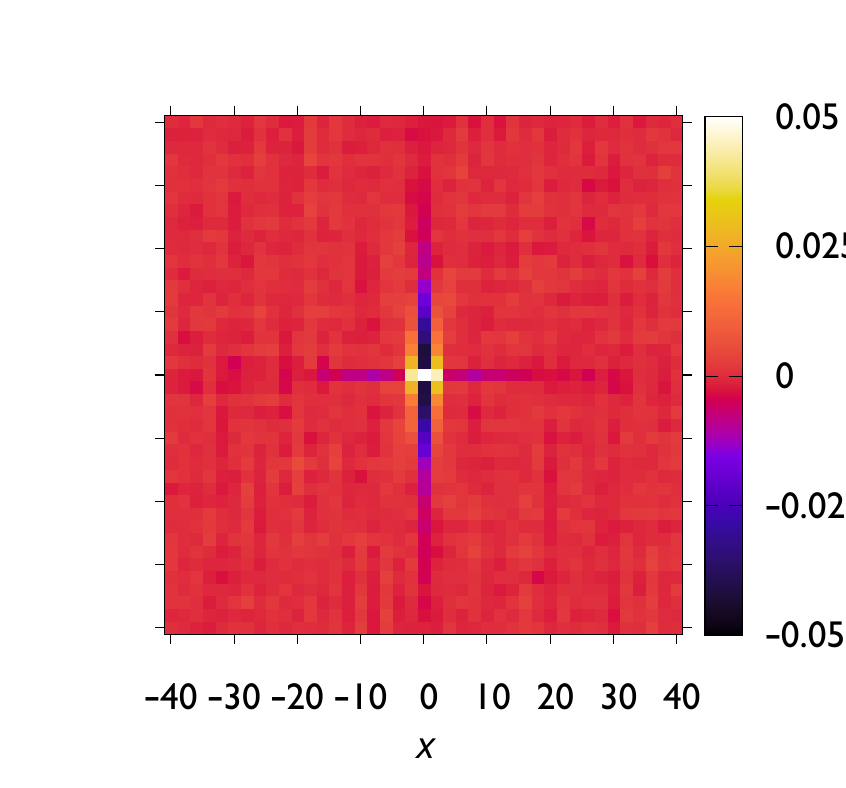}
\vspace*{-2mm}

\vspace*{-2mm}
\begin{picture}(0,0)
\put(-10,28){\rotatebox{90}{\sffamily{$\gamma=1$}}}
\end{picture}
\hspace*{2mm}
\includegraphics[angle=0, height=3.1cm, trim=25mm 0 22mm 0]{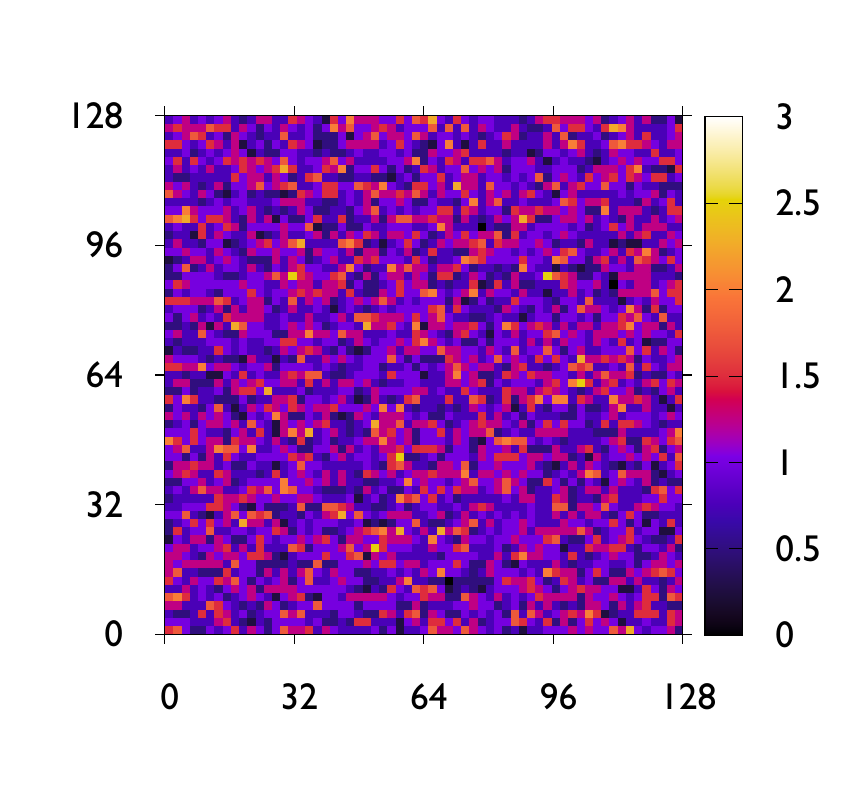}
\includegraphics[angle=0, height=3.1cm, trim=5mm 0 0 0]{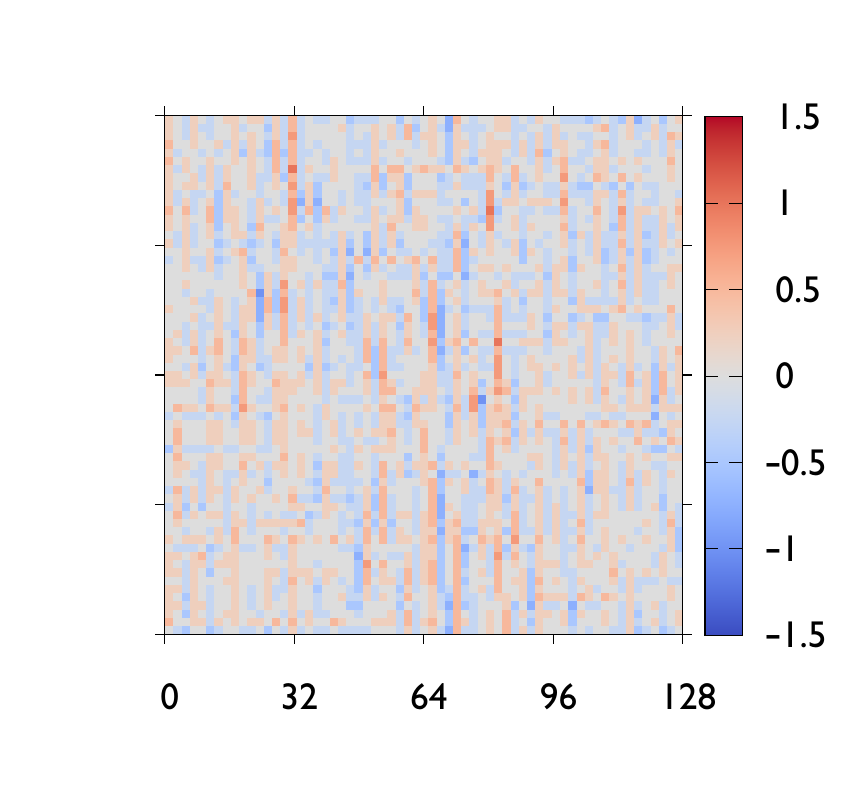}
\includegraphics[angle=0, height=3.1cm, trim=0mm 0 8mm 0]{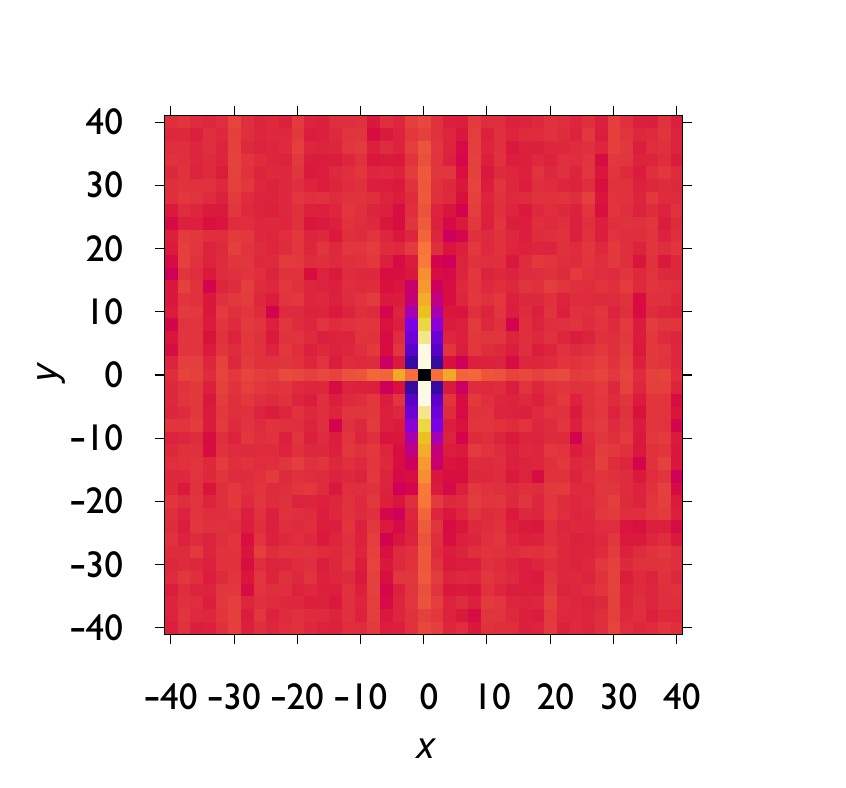}
\includegraphics[angle=0, height=3.1cm, trim=35mm 0 8mm 0]{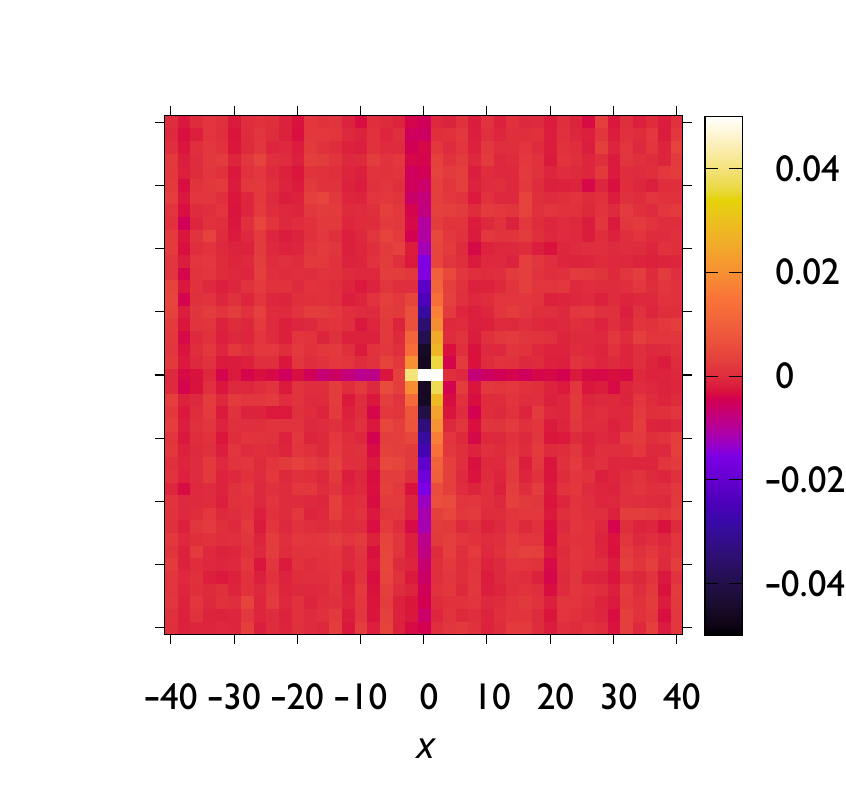}
\vspace*{-2mm}

\vspace*{-2mm}
\begin{picture}(0,0)
\put(-10,28){\rotatebox{90}{\sffamily{$\gamma=16$}}}
\end{picture}
\hspace*{2mm}
\includegraphics[angle=0, height=3.1cm, trim=25mm 0 22mm 0]{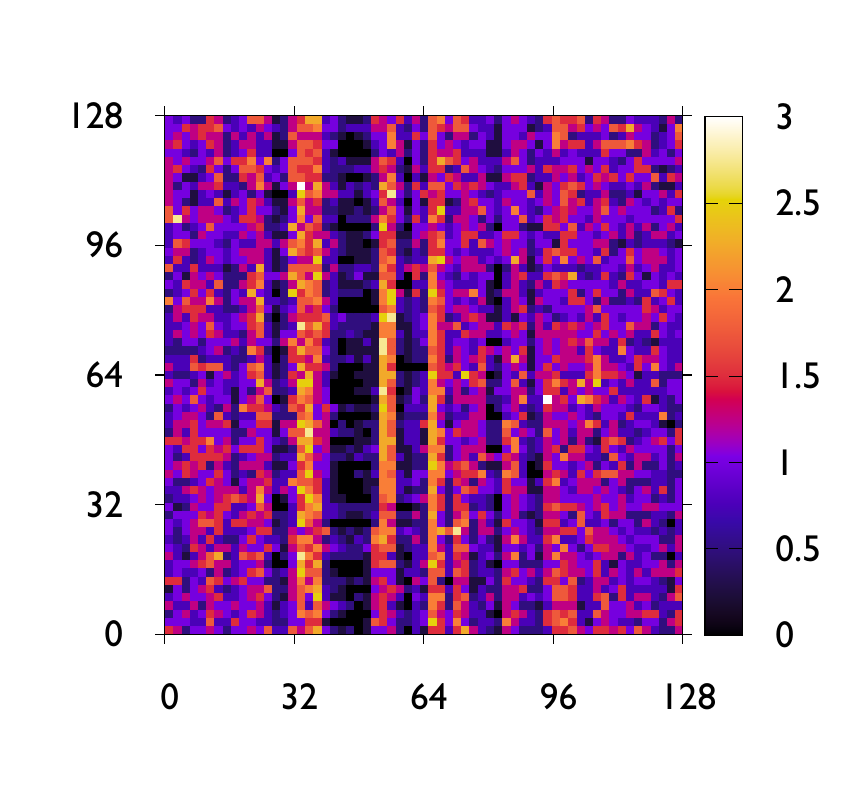}
\includegraphics[angle=0, height=3.1cm, trim=5mm 0 0 0]{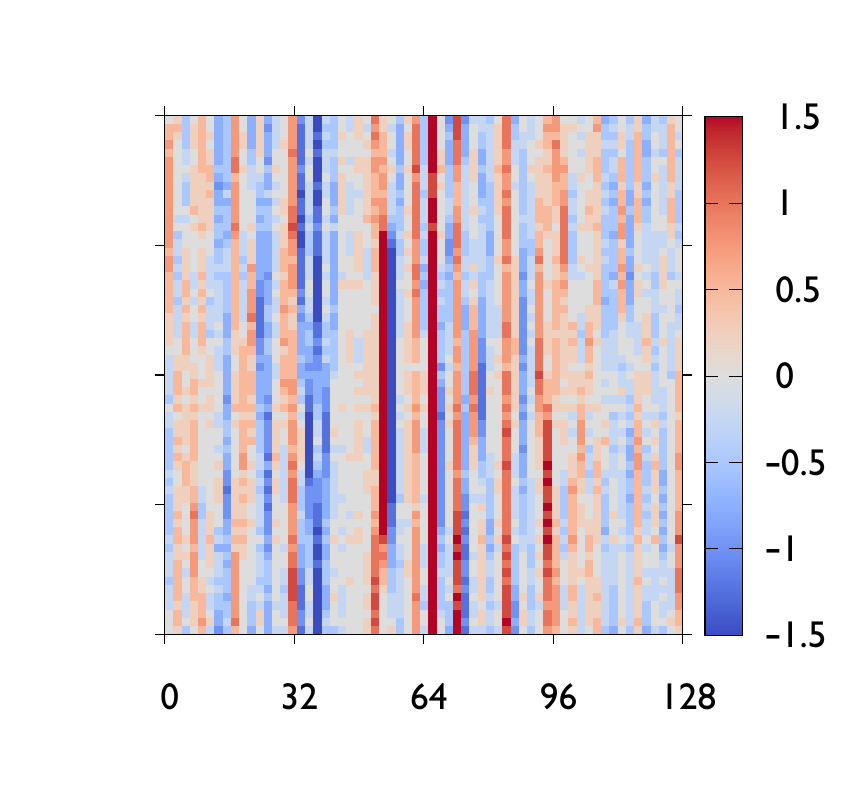}
\includegraphics[angle=0, height=3.1cm, trim=0mm 0 8mm 0]{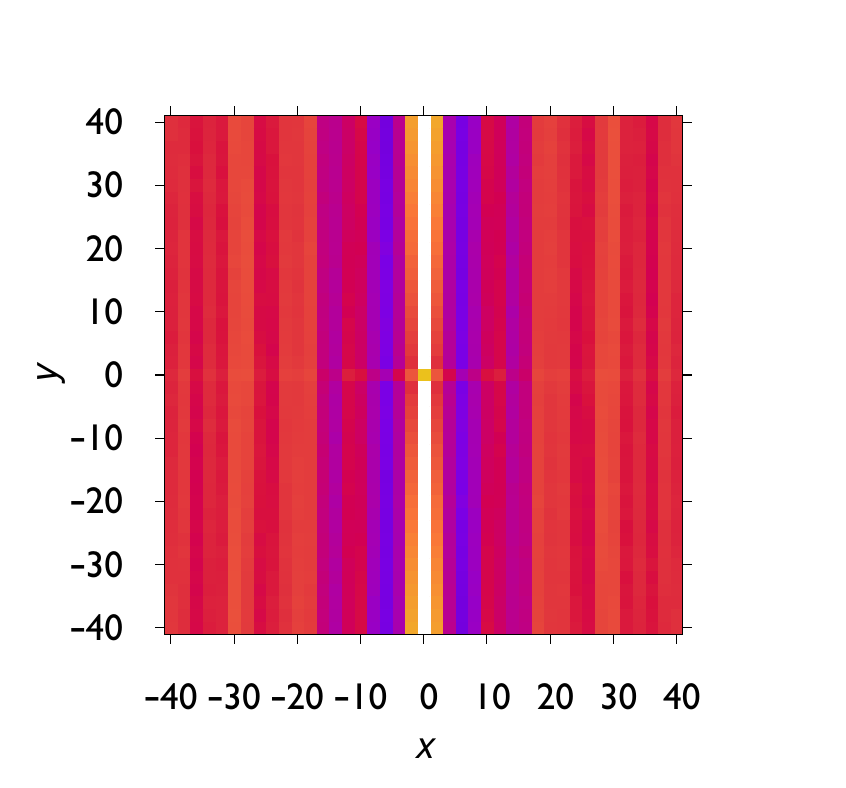}
\includegraphics[angle=0, height=3.1cm, trim=35mm 0 8mm 0]{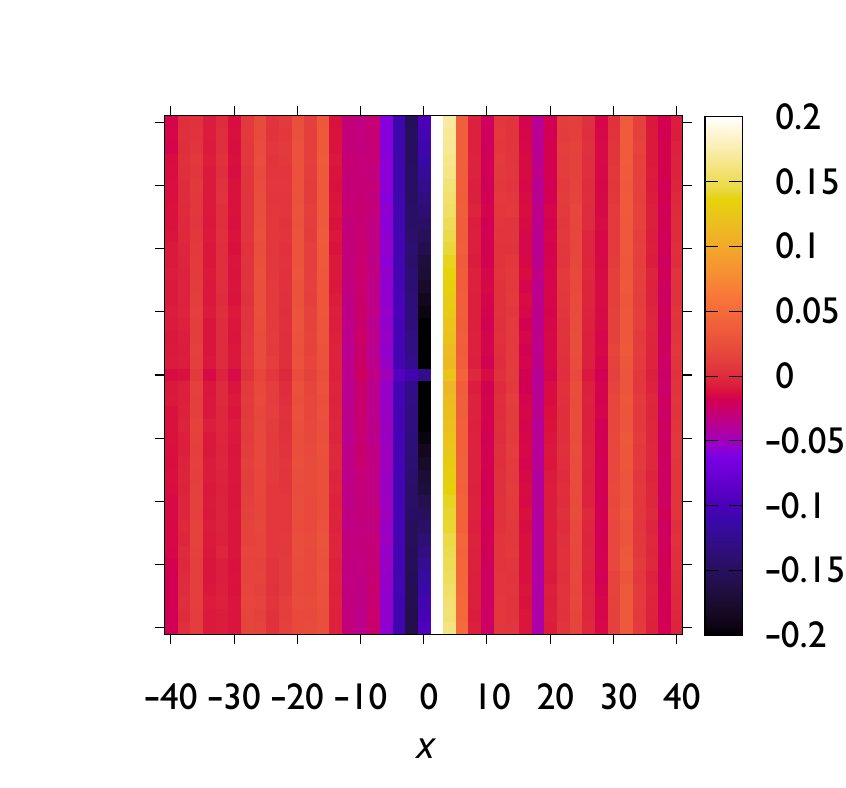}
\vspace*{-2mm}

\vspace*{-2mm}
\begin{picture}(0,0)
\put(-10,28){\rotatebox{90}{\sffamily{$\gamma=128$}}}
\end{picture}
\hspace*{2mm}
\includegraphics[angle=0, height=3.1cm, trim=25mm 0 22mm 0]{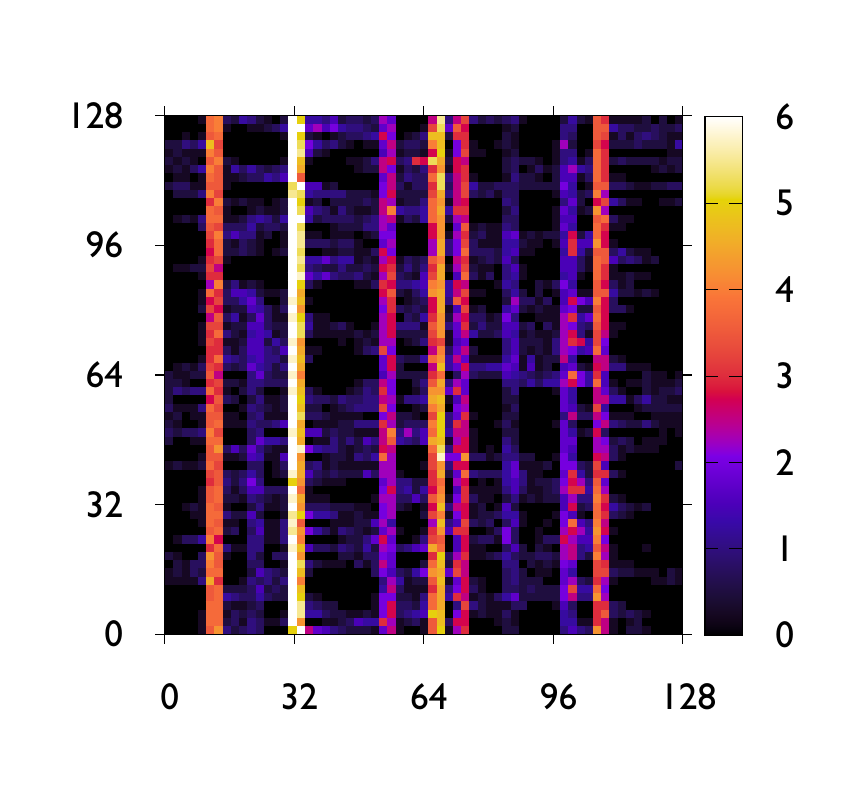}
\includegraphics[angle=0, height=3.1cm, trim=5mm 0 0 0]{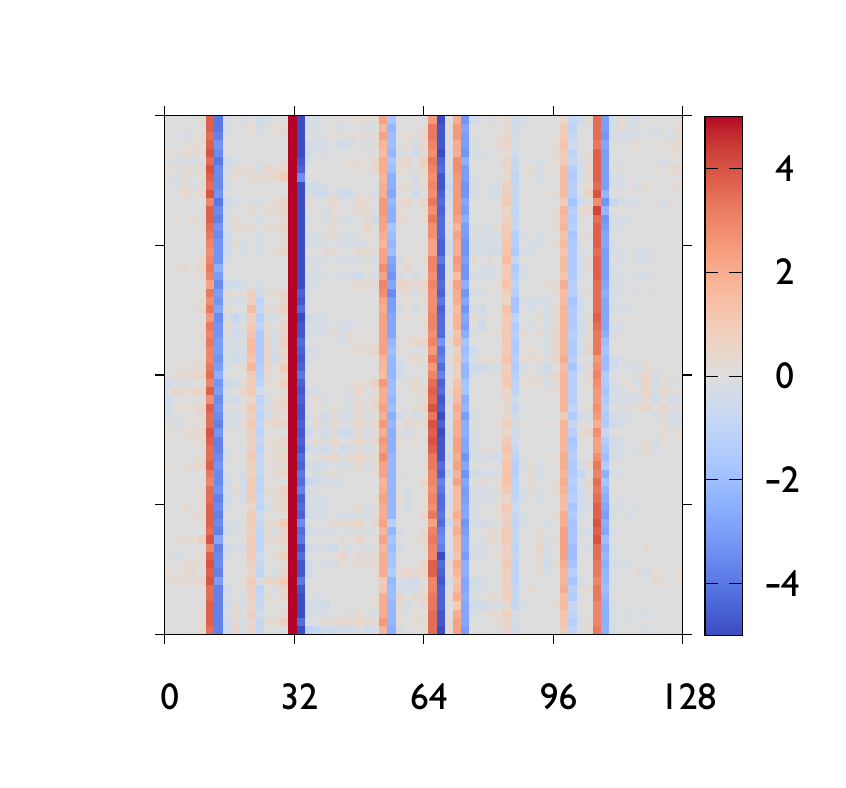}
\includegraphics[angle=0, height=3.1cm, trim=0mm 0 8mm 0]{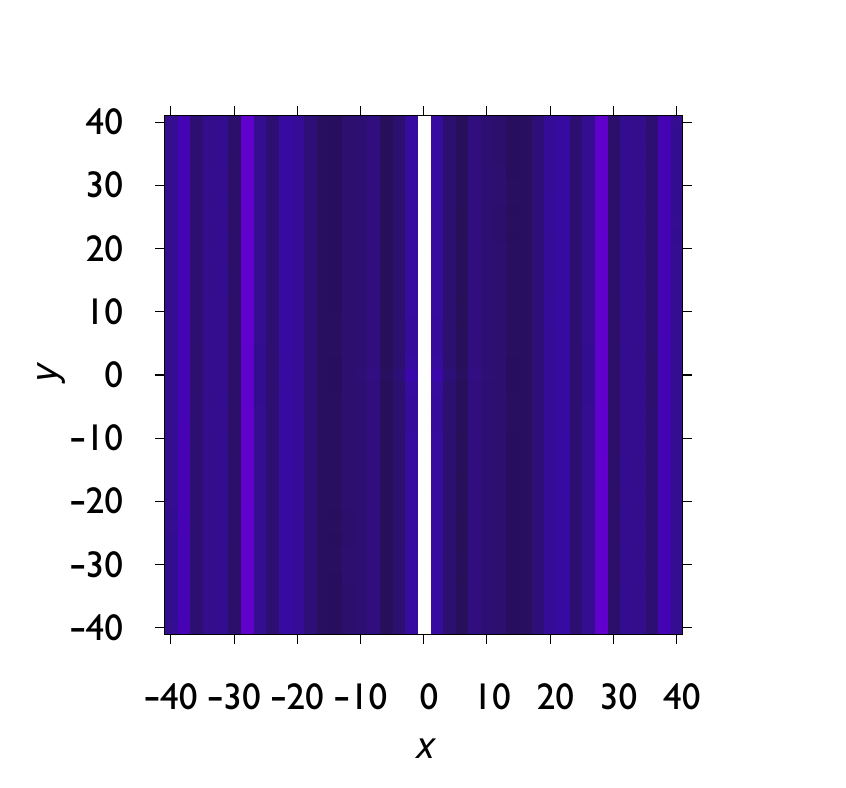}
\includegraphics[angle=0, height=3.1cm, trim=35mm 0 8mm 0]{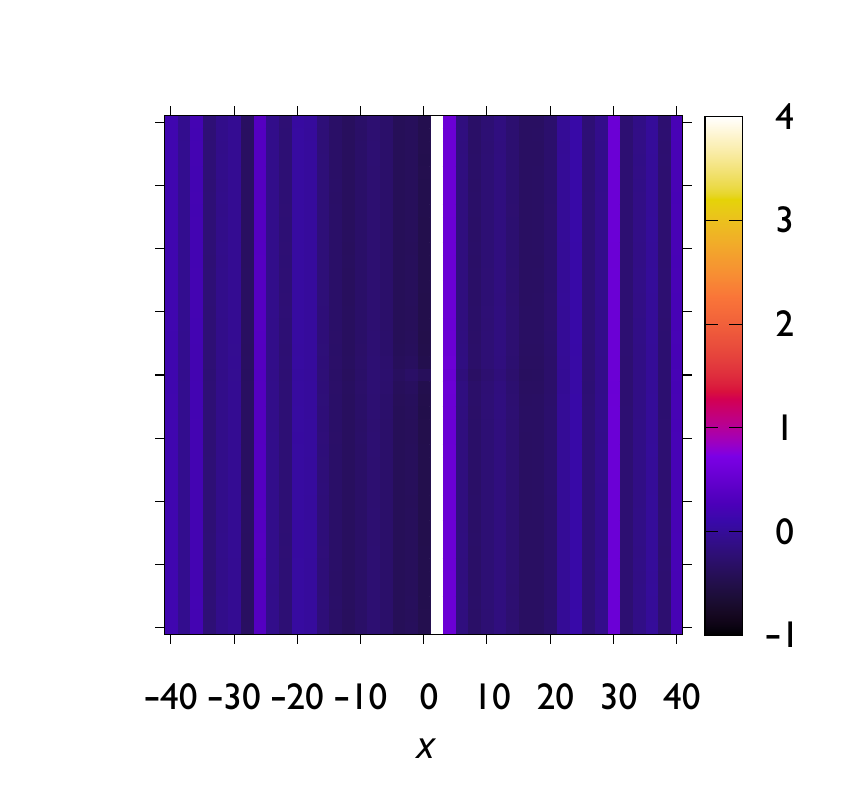}
\vspace*{-2mm}

%
%\hspace*{-10mm}
%\begin{picture}(0,0)
%\put(-12,100){\sffamily{(b)}}
%\end{picture}
%\includegraphics[angle=0, height=4cm, trim=4mm 0 8mm 0]{Figures/stress_strain/stress_strain_comparison.pdf}
\end{center}
\vspace*{-4mm}
\caption{Dislocation pattern evolution in SCDD simulations with $D=A=0.25$, $\alpha=1.0$ and $a=2$. Left two columns: Total and GND density maps obtained at different strain $\gamma$ values. Right two columns: Same sign and opposite sign spatial correlation functions ($d_{++}$ and $d_{+-}$, respectively) of dislocation densities.
\vspace*{-0.2cm}
\label{fig:pattern_evol_continuum}}
\end{figure*}

The numerical implementation is based on the phase-field functional:
\begin{equation}
P[\rho, \kappa] = E_\text{el} + \int \left[ A \rho \ln\left( \frac{\rho}{\rho_0} \right) + \frac{D(p)}{2} \frac{\kappa^2}{\rho}\right] \mathrm d^2 r,
\label{eq:plastic_potential}
\end{equation}
where $E_\text{el}$ is the mean-field stored elastic energy of the system (measured in units of $Gb^2$) \cite{groma_dislocation_2016}, and $\rho_0$ is a constant that does not appear in the evolution equations. It was shown before that Eqs.~(\ref{eq:rhod2plus},\ref{eq:rhod2minus}) can be derived from Eq.~(\ref{eq:plastic_potential}) assuming that $|\kappa| \ll \rho$ and that $P$ can only decrease during the evolution of the system \cite{groma_dislocation_2016}. In the present implementation, densities are discretized on a regular grid of cell size $a$, and the flow stress $\tau_f$ is replaced by a local stochastic variable (representing the fluctuations of the underlying dislocation microstructure at every cell). For the distribution of the flow stress, in accordance with recent DDD results, a Weibull distribution is used with shape parameter 1.4 and scale parameter $\alpha$ \cite{ispanovity_role_2017}. We apply extremal dynamics: At every timestep, dislocation activity takes place at the site where decrease in $P$ is the largest and this consists of a quantum of dislocation flux $\Delta \rho=a^{-2}$ (of either positive or negative dislocations) flowing through the cell boundary. If no such cell exists, external stress $\tau_\text{ext}$ is increased until one cell is triggered. Further details of the implementation are summarized in the Supplemental Material. In the rest of this paper  this model is refered to as stochastic continuum dislocation dynamics (SCDD).

The continuum theory does not yield exact values for the parameters $\alpha$, $D$, and $A$; one must, therefore, consider them as fitting parameters. The results of DDD simulations summarized above, however, give insight on possible values. As seen in Fig.~\ref{fig:pattern_evol_ddd} the strongest possible dislocation configuration is the dipolar wall structure. According to Fig.~\ref{fig:sketch} this might imply the necessity of the back-stress term $\tau_b$ and that $D<0$. However, according to the linear stability analysis of Eqs.~(\ref{eq:rhod2plus},\ref{eq:rhod2minus}), in the absence of stochastic terms, which is discussed in Ref.~\cite{groma_dislocation_2016}, one may conclude that if $D=A=0$ or either $D$ or $A$ is negative, then all perturbations are unstable, and the fastest growing perturbation is seemingly the largest possible wave-vector $k_x=2\pi/|b|$ where $|b|$ is the minimum-possible Burgers vector magnitude. This implies that in such cases the wavelength of the instability approaches the lattice scale, and one would expect a dislocation pattern where positive/negative dislocation walls arrange in a $+-+-$ type instability. In the following, that is exactly what we observe. Moreover, if $A>0$ and $D>0$, resulting that the phase field functional given by Eq. (\ref{eq:plastic_potential}) is convex,  then the linear stability analysis of Ref.~\cite{groma_dislocation_2016} concludes that there is an emergent wavelength selection for a periodic pattern that should appear. However, in the presence of quenched stochastic terms, such as the ones considered both in the presented DDD and SCDD models, it is expected that such instabilities would be suppressed by the quenched disorder, which typically takes the form of a stochastic flow stress term. The inclusion of a stochastic flow stress term (such as the one in this work) is critical to suppress non-linear dynamical instabilities that are tied to the particular deterministic dynamical equations and thus, non-generic, and probably this difference may explain particularly different conclusions of our work compared to prior efforts on analogous questions \cite{PhysRevB.98.054110}.

\begin{figure}[!ht]
\begin{picture}(0,0)
\put(5,125){\sffamily{(a)}}
\end{picture}
\includegraphics[angle=0, width=6.5cm]{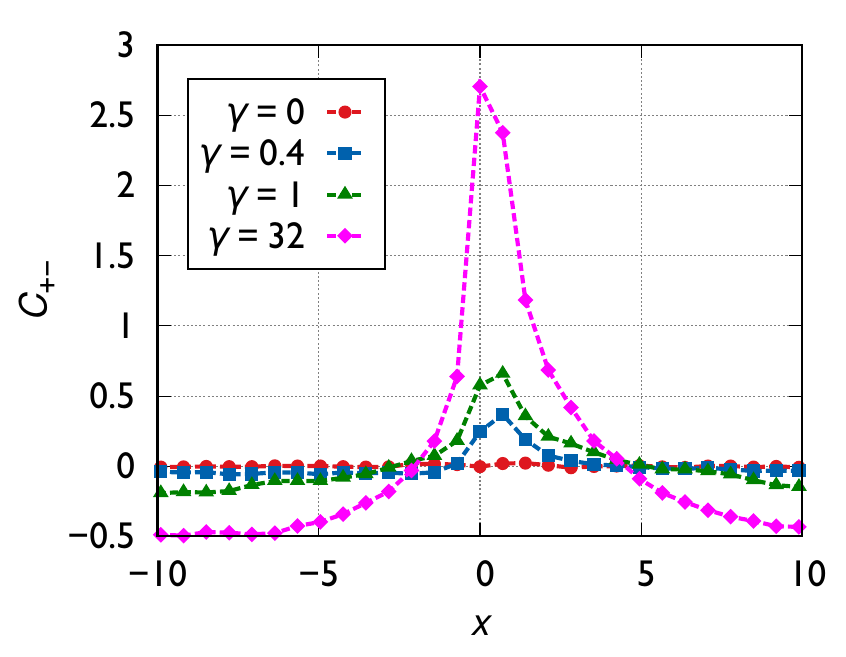}

\begin{picture}(0,0)
\put(5,125){\sffamily{(b)}}
\end{picture}
\includegraphics[angle=0, width=6.5cm]{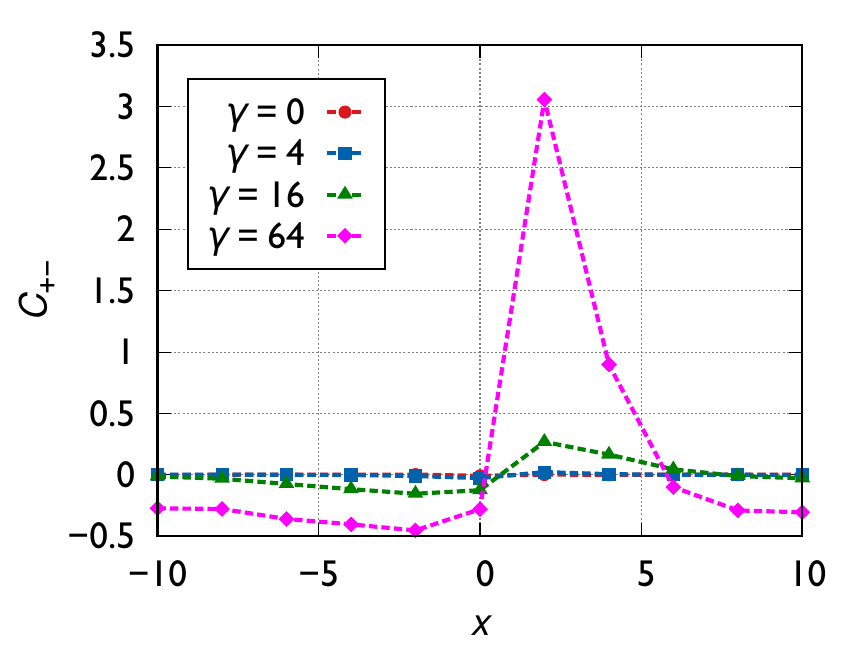}
\caption{Correlation functions averaged along the DDW direction $C_{+-}$ for the two applied models: (a) DDD and (b) SCDD. Notice the asymmetry developing upon increasing strain. The simulation parameters are the same as in Figs.~\ref{fig:bck} and \ref{fig:pattern_evol_continuum}.
\label{fig:cpm}
}
\end{figure}

In accordance with our 2D-DDD simulations, at $t=0$ a random pattern of $\rho_+$ and $\rho_-$ is assumed, and initially a relaxation step is performed at $\tau_\text{ext}=0$. Then, the external stress $\tau_\text{ext}$ is quasi-statically increased. The two left columns of Fig.~\ref{fig:pattern_evol_continuum} depict this evolution in terms of the total and GND densities ($\rho$ and $\kappa$, respectively) for a given parameter set ($D=0.25$, $A=0.25$, $\alpha=1.0$, $a=2$). Analogous behavior is obtained between DDD and SCDD in the evolution of both total and GND profiles: short vertical DDWs form in the initial phases of deformation which then merge and extend in the vertical direction upon increasing strain. To descibe the patterns in more detail, the correlation functions $d_{++}$ and $d_{+-}$ were also calculated and can be seen in the right two columns of Fig.~\ref{fig:pattern_evol_continuum}. Although fine details of the correlation functions cannot be reproduced with a continuum method defined on a coarse mesh, the evolution of these functions is remarkably similar to the DDD case (seen in Fig.~\ref{fig:pattern_evol_ddd}), however, pattern development is somewhat delayed in the SCDD (also note the difference in the spatial scale between DDD and SCDD results). In particular, at strain $\gamma=0$ the configuration consists of short vertical walls of same sign dislocations and short dislocation dipoles of opposite sign having an angle of 45 degrees with the $x$ axis. As the stress and strain increases the vertical walls become more extended and strong asymmetry develops in the $d_{+-}$ correlation function signaling the presence of DDWs. To quantitatively compare the polarization due to the asymmetry in $d_{+-}$ between DDD and SCDD we consider the spatial average of $d_{+-}$ along the vertical direction as
\begin{equation}
	C_{+-}(x) = (1/L) \int d_{+-}(x,y)  \text{d} y
\end{equation}
which measures the polarization of individual configurations. Indeed Fig.~\ref{fig:cpm} shows that a strong asymmetry emerges upon plastic deformation for both models. In addition, a quick, exponential decay follows the peak in the $x>0$ domain, with a characteristic distance of approx.~5 average dislocation spacings. This length scale can be identified with the characteristic width of the DDWs. We identify this asymmetry as the most basic origin of the Bauschinger effect~\cite{abel_bauschinger_1972} as it will be discussed below.

\begin{figure*}[!t]
\begin{center}
\vspace*{1.0cm}
\begin{picture}(0,0)
%\put(-3,95){\sffamily{(a)}}
\put(-35,-10){\rotatebox{90}{\parbox{4cm}{\sffamily{$D=0.25$,\\$A=0.25,$\\ $\alpha=1.0$}}}}
\put(8,95){\parbox{1in}{\sffamily{Total density\\($\rho$)}}}
\put(90,95){\parbox{1in}{\sffamily{GND density\\($\kappa$)}}}
\put(185,95){\parbox{1in}{\sffamily{Same sign correlation function ($d_{++}$)}}}
\put(258,95){\parbox{1in}{\sffamily{Opposite sign correlation function ($d_{+-}$)}}}
\end{picture}
\hspace*{2mm}
\includegraphics[angle=0, height=3.1cm, trim=25mm 0 22mm 0]{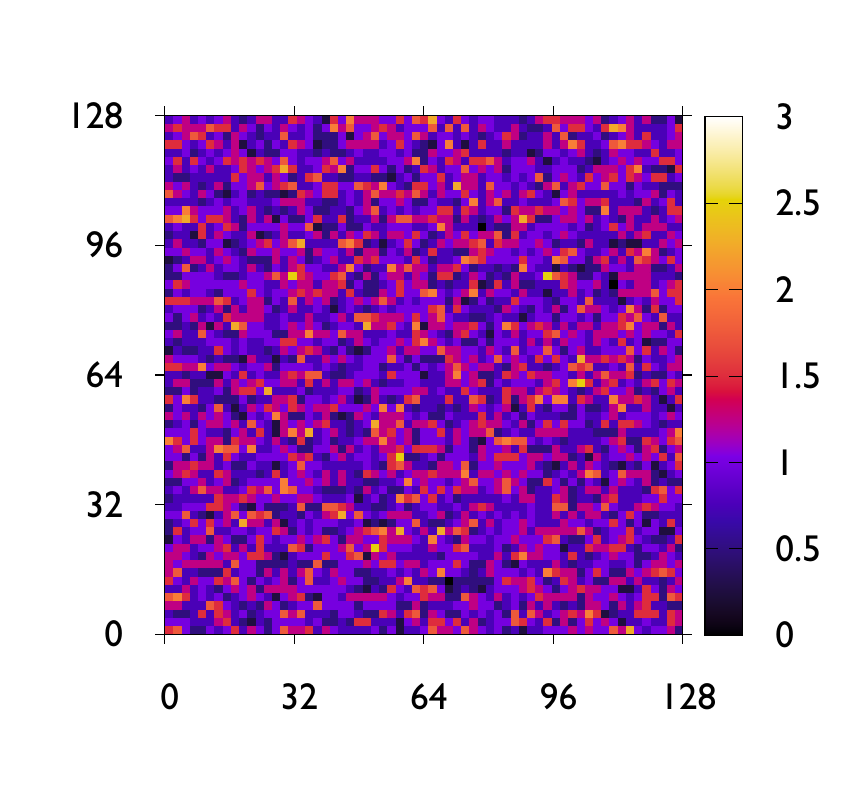}
\includegraphics[angle=0, height=3.1cm, trim=5mm 0 0 0]{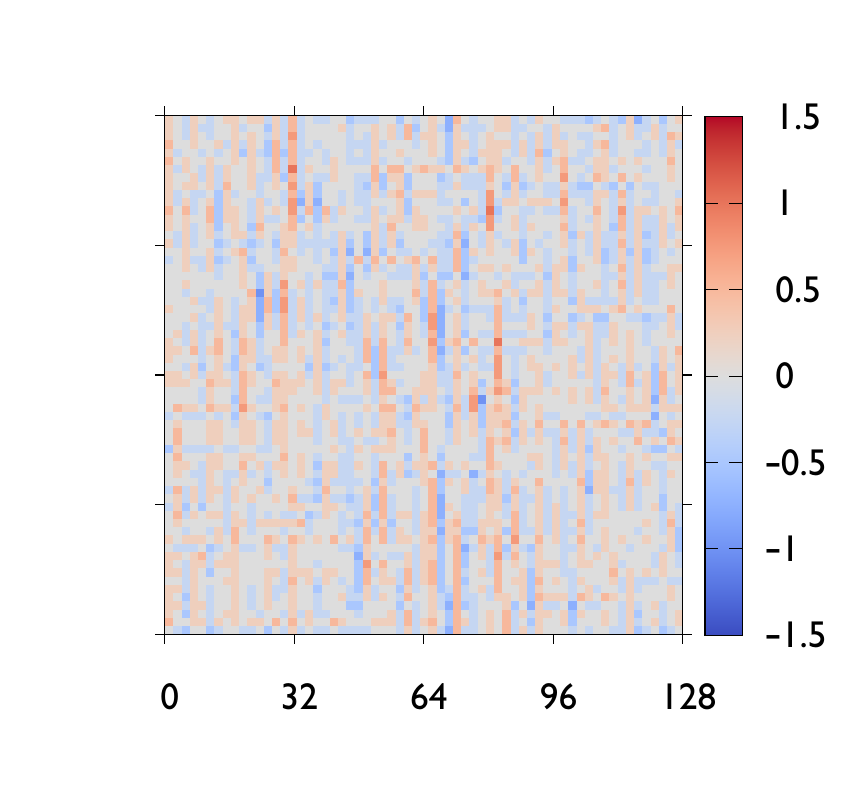}
\includegraphics[angle=0, height=3.1cm, trim=0mm 0 8mm 0]{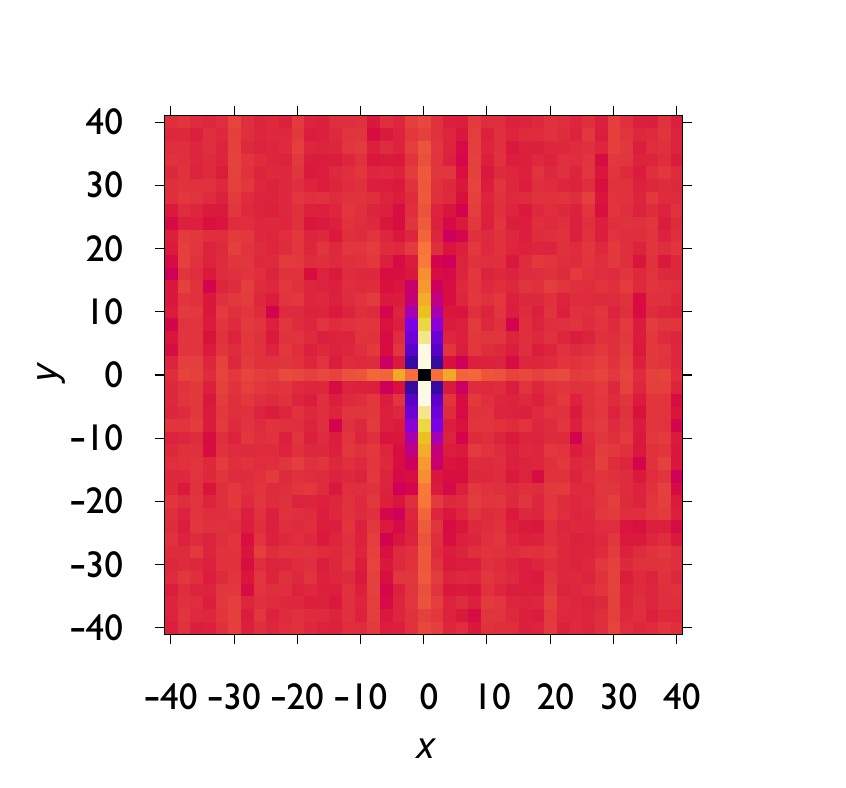}
\includegraphics[angle=0, height=3.1cm, trim=35mm 0 8mm 0]{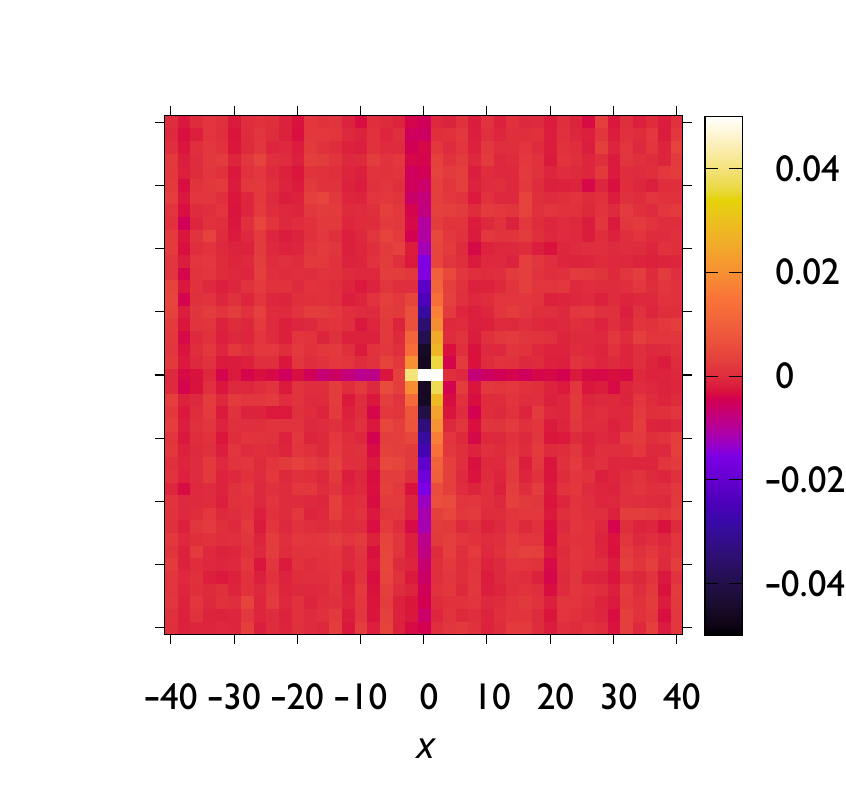}
\vspace*{-2mm}

\vspace*{-2mm}
\begin{picture}(0,0)
\put(-35,-10){\rotatebox{90}{\parbox{4cm}{\sffamily{$D=0.0$,\\$A=0.0,$\\ $\alpha=1.0$}}}}
\end{picture}
\hspace*{2mm}
\includegraphics[angle=0, height=3.1cm, trim=25mm 0 22mm 0]{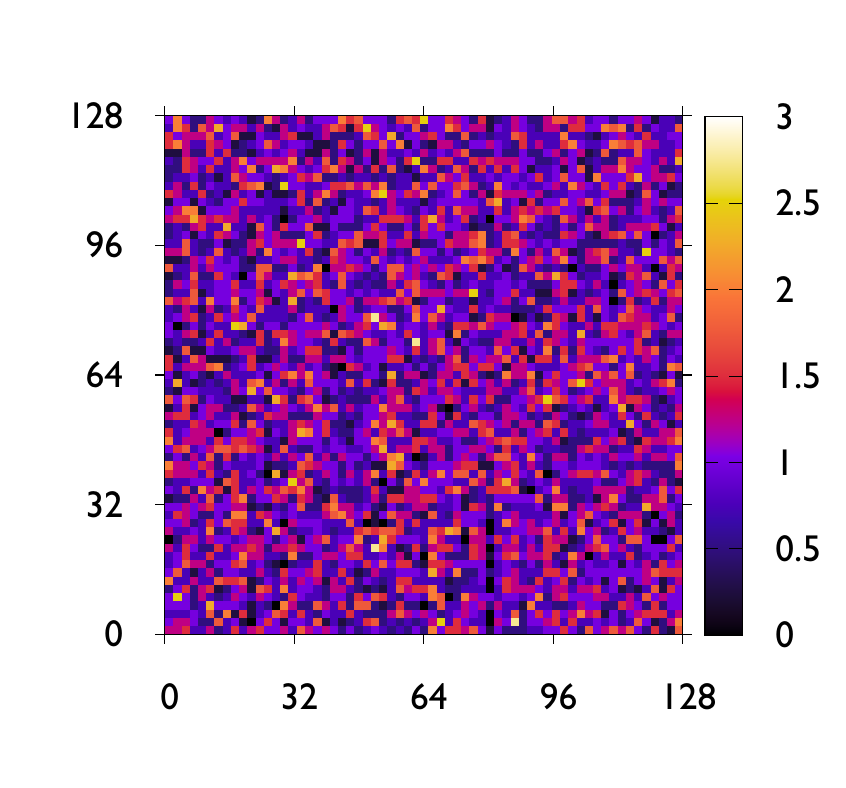}
\includegraphics[angle=0, height=3.1cm, trim=5mm 0 0 0]{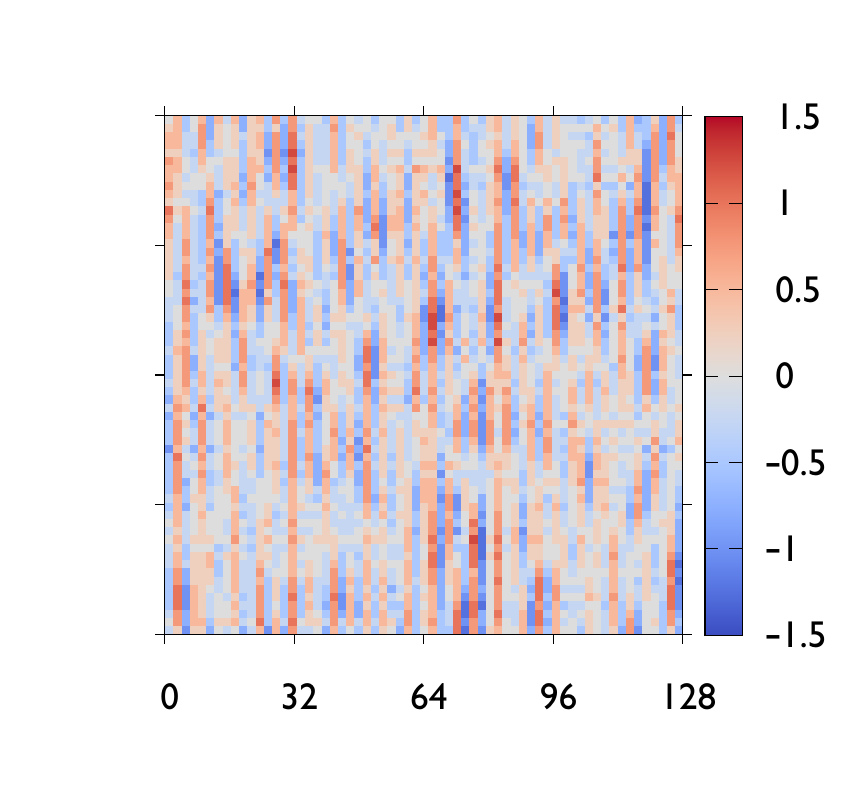}
\includegraphics[angle=0, height=3.1cm, trim=0mm 0 8mm 0]{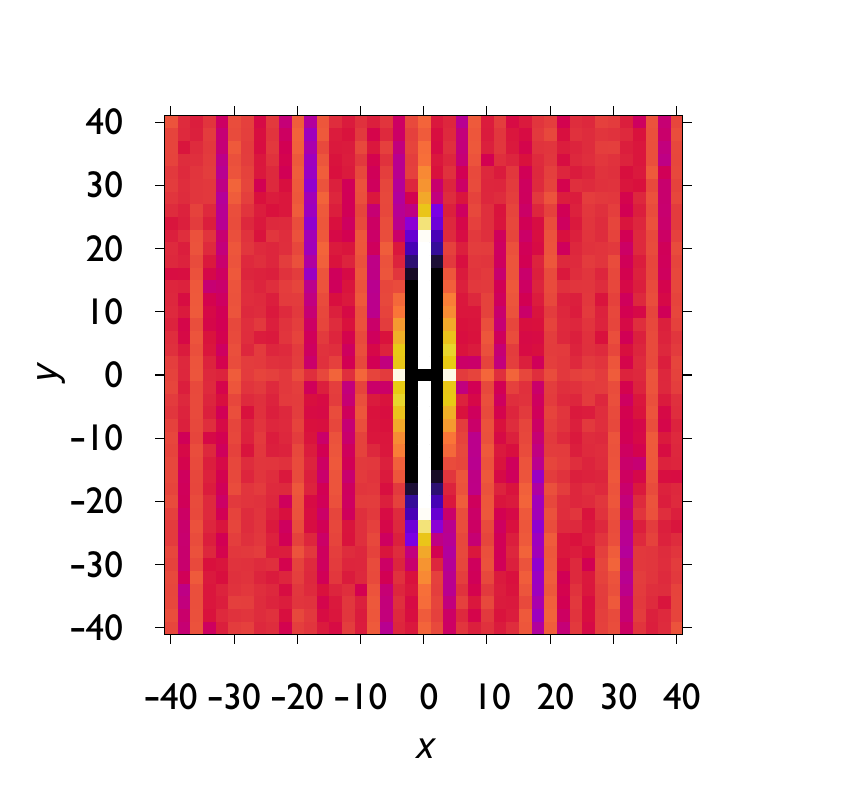}
\includegraphics[angle=0, height=3.1cm, trim=35mm 0 8mm 0]{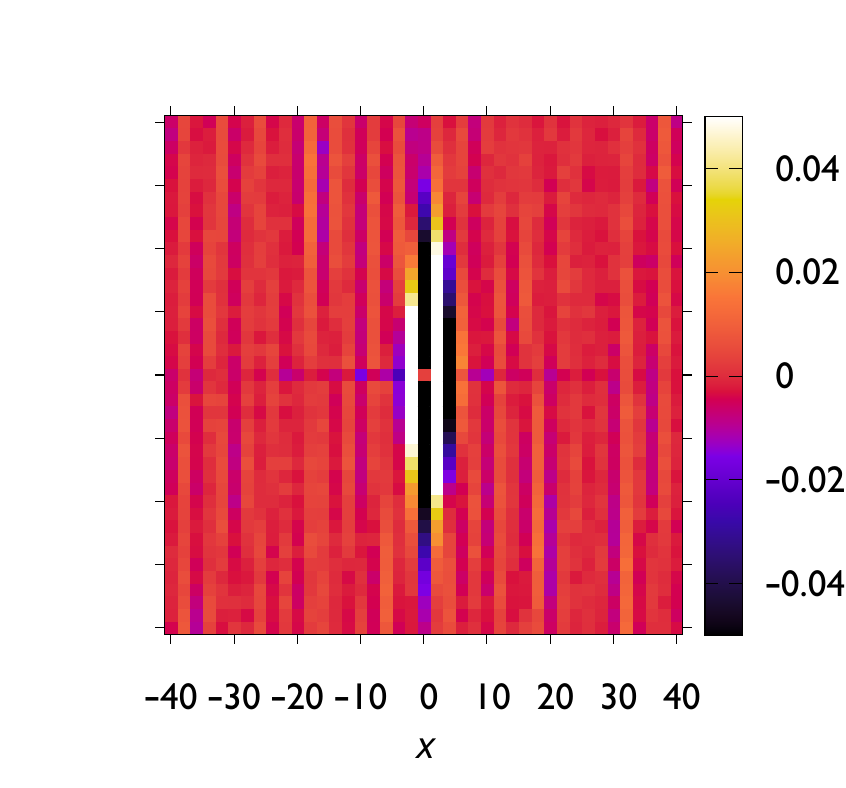}
\vspace*{-2mm}

\vspace*{-2mm}
\begin{picture}(0,0)
\put(-35,-10){\rotatebox{90}{\parbox{4cm}{\sffamily{$D=-0.25$,\\$A=0.25,$\\ $\alpha=1.0$}}}}
\end{picture}
\hspace*{2mm}
\includegraphics[angle=0, height=3.1cm, trim=25mm 0 22mm 0]{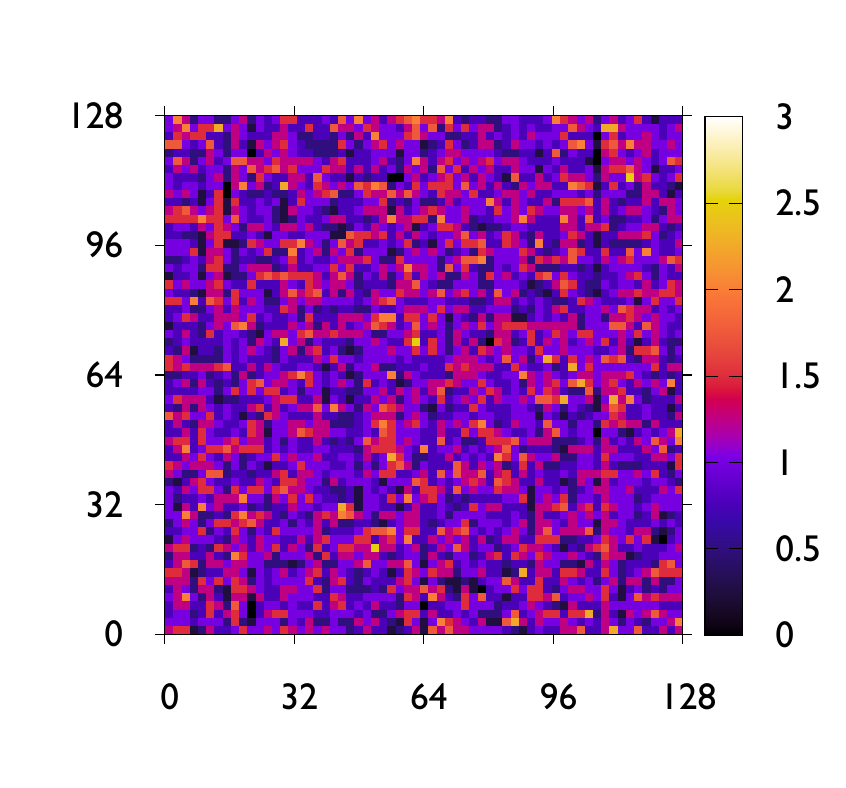}
\includegraphics[angle=0, height=3.1cm, trim=5mm 0 0 0]{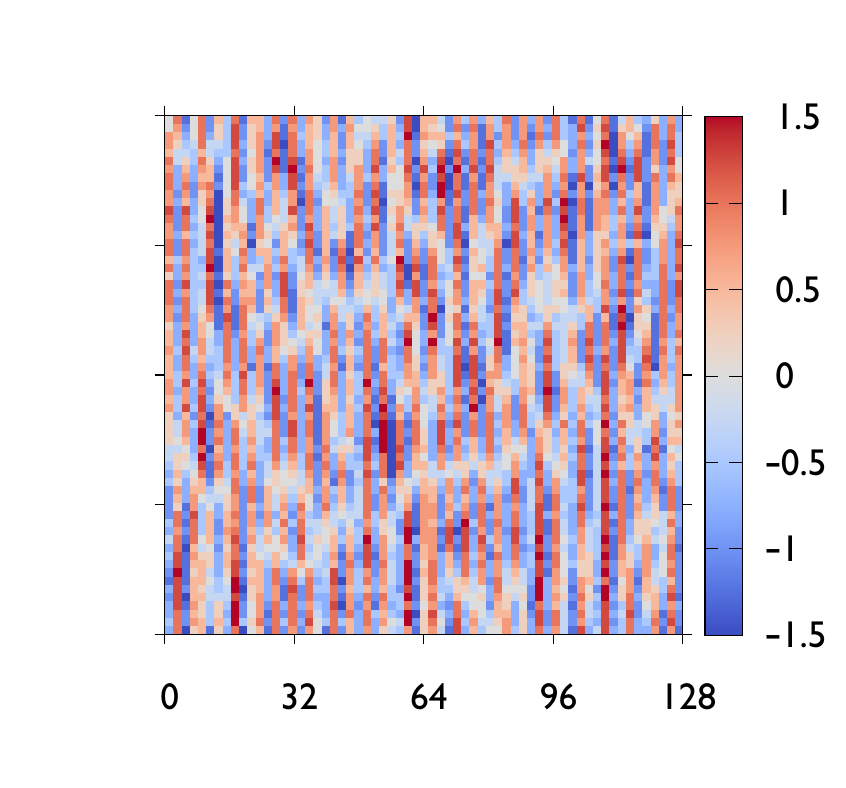}
\includegraphics[angle=0, height=3.1cm, trim=0mm 0 8mm 0]{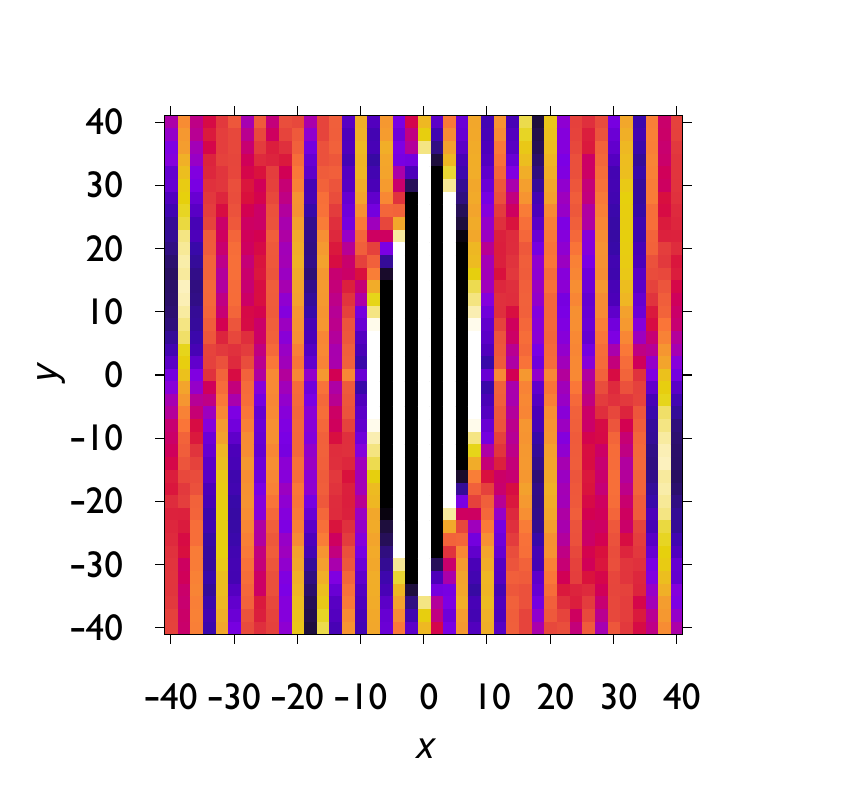}
\includegraphics[angle=0, height=3.1cm, trim=35mm 0 8mm 0]{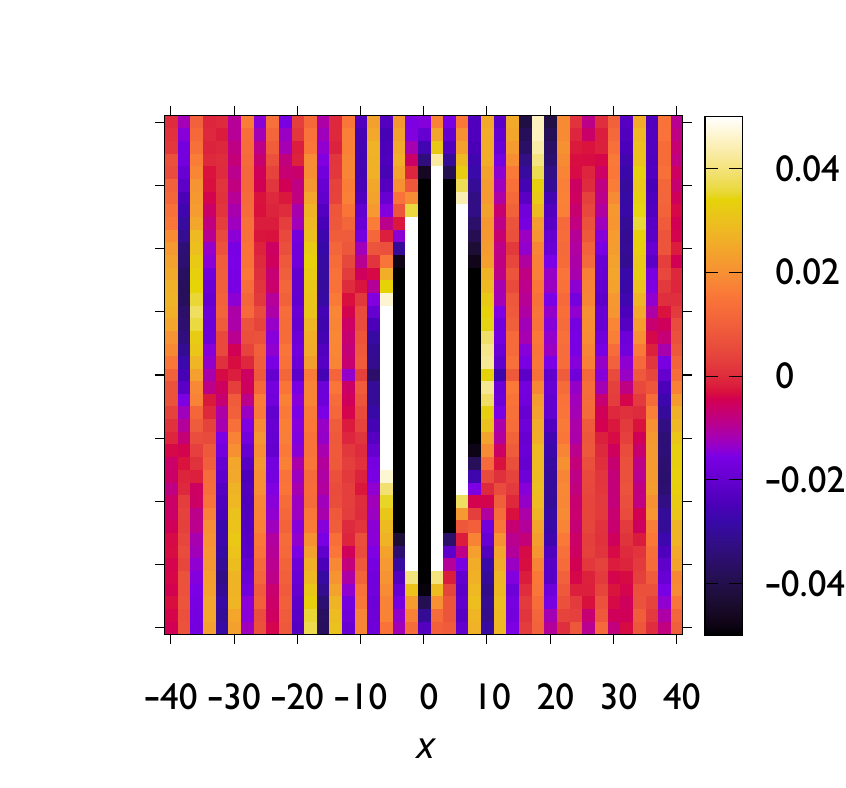}
\vspace*{-2mm}

\vspace*{-2mm}
\begin{picture}(0,0)
\put(-35,-10){\rotatebox{90}{\parbox{4cm}{\sffamily{$D=0.25$,\\$A=1.0,$\\ $\alpha=1.0$}}}}
\end{picture}
\hspace*{2mm}
\includegraphics[angle=0, height=3.1cm, trim=25mm 0 22mm 0]{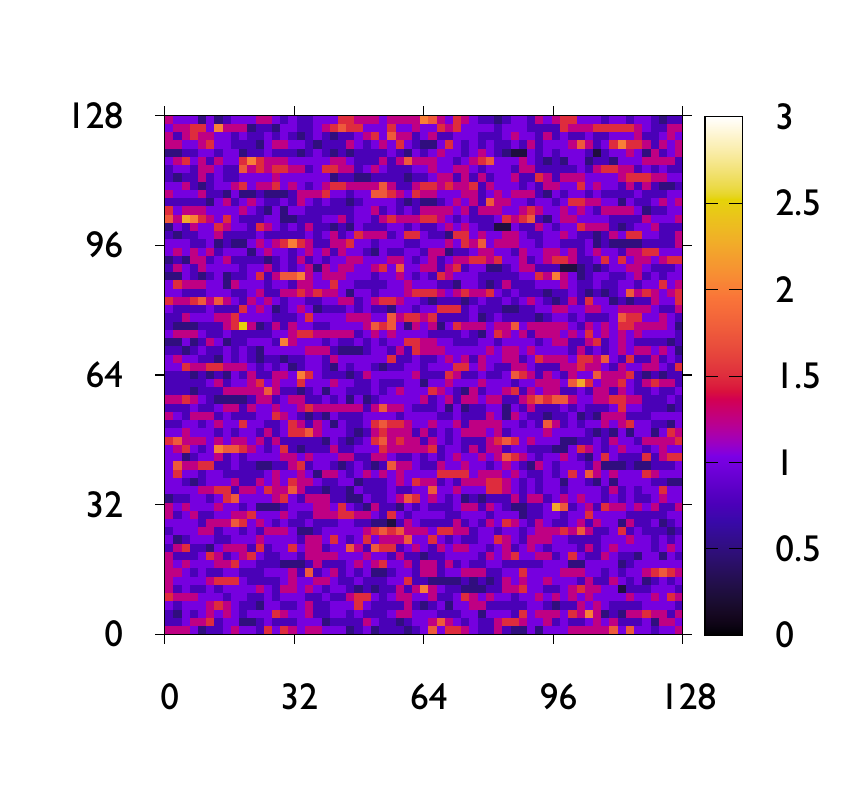}
\includegraphics[angle=0, height=3.1cm, trim=5mm 0 0 0]{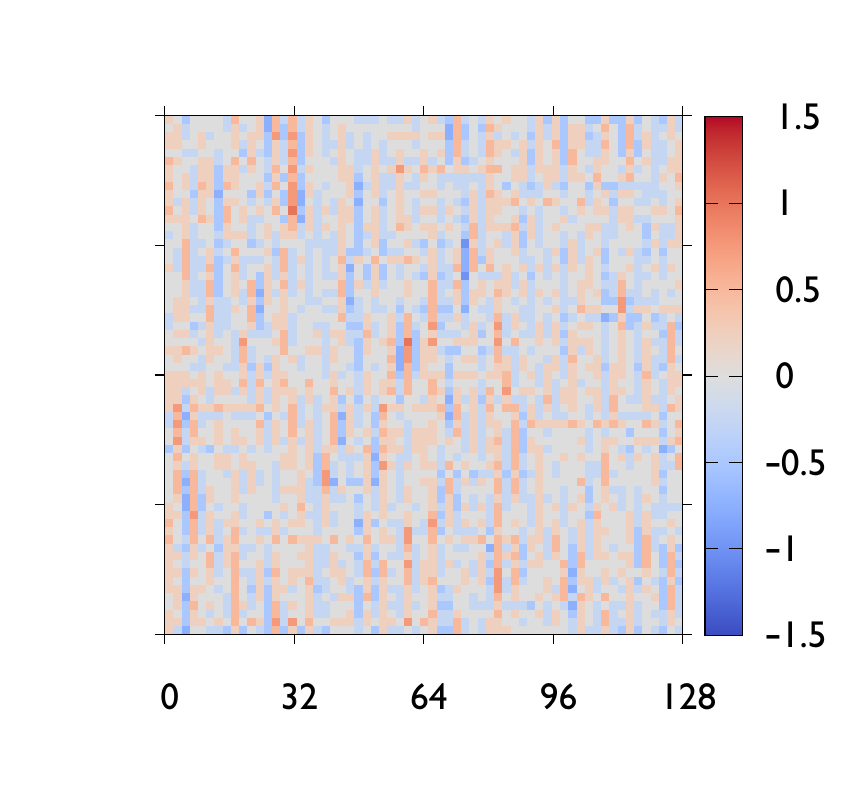}
\includegraphics[angle=0, height=3.1cm, trim=0mm 0 8mm 0]{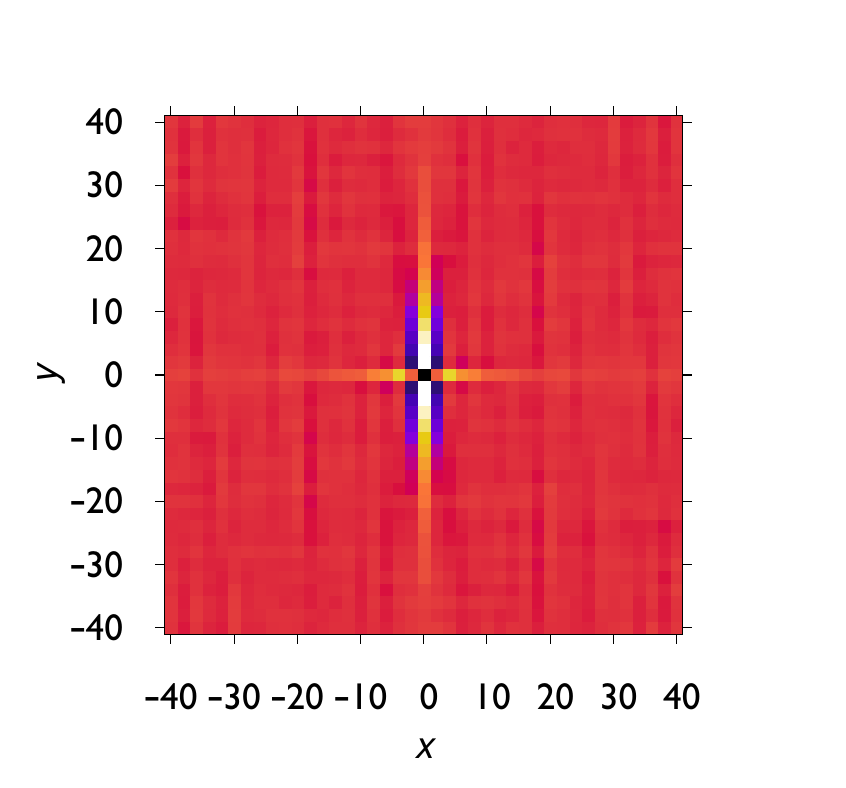}
\includegraphics[angle=0, height=3.1cm, trim=35mm 0 8mm 0]{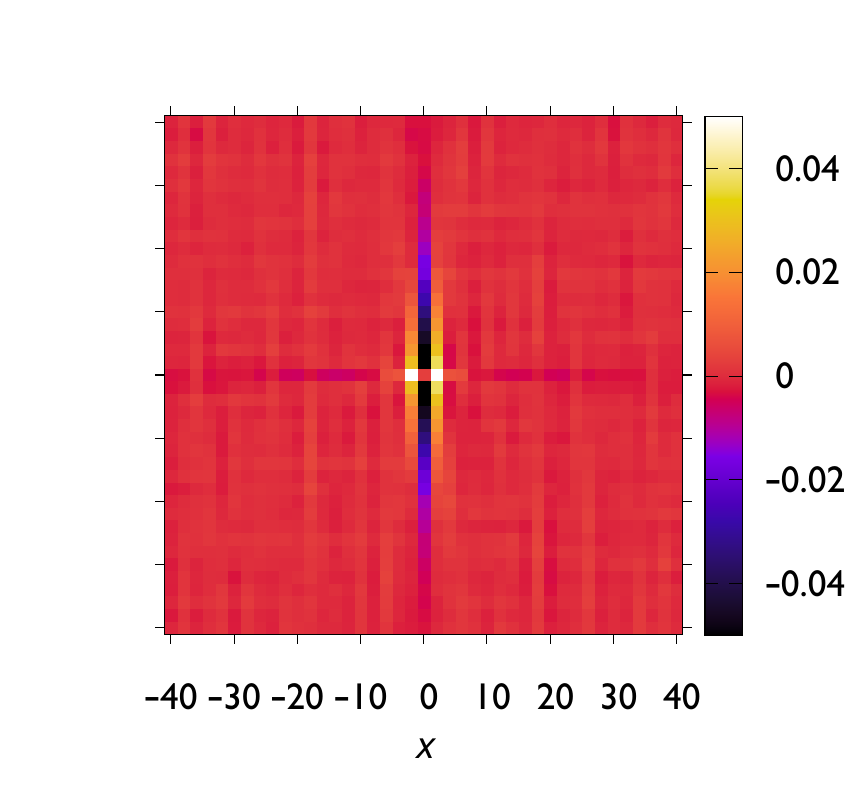}
\vspace*{-2mm}

\vspace*{-2mm}
\begin{picture}(0,0)
\put(-35,-10){\rotatebox{90}{\parbox{4cm}{\sffamily{$D=0.25$,\\$A=0.25,$\\ $\alpha=0.25$}}}}
\end{picture}
\hspace*{2mm}
\includegraphics[angle=0, height=3.1cm, trim=25mm 0 22mm 0]{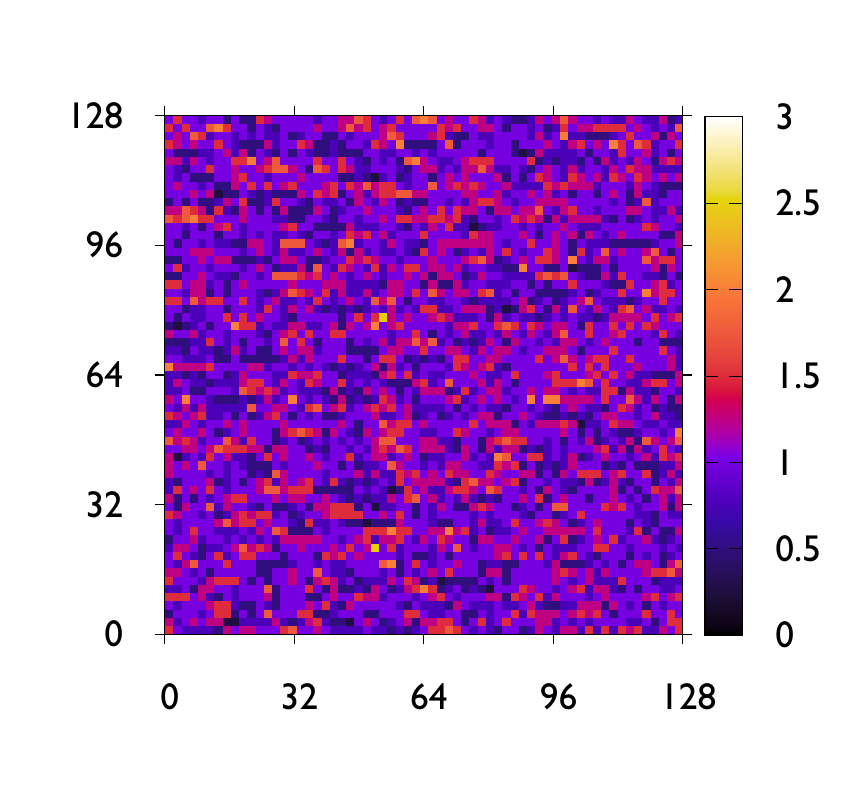}
\includegraphics[angle=0, height=3.1cm, trim=5mm 0 0 0]{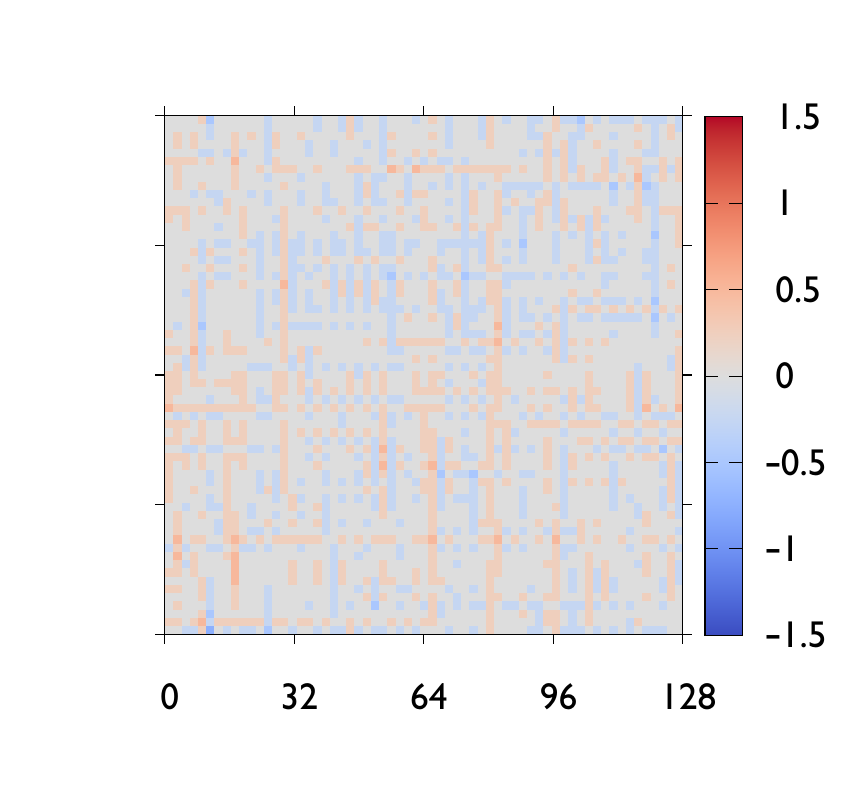}
\includegraphics[angle=0, height=3.1cm, trim=0mm 0 8mm 0]{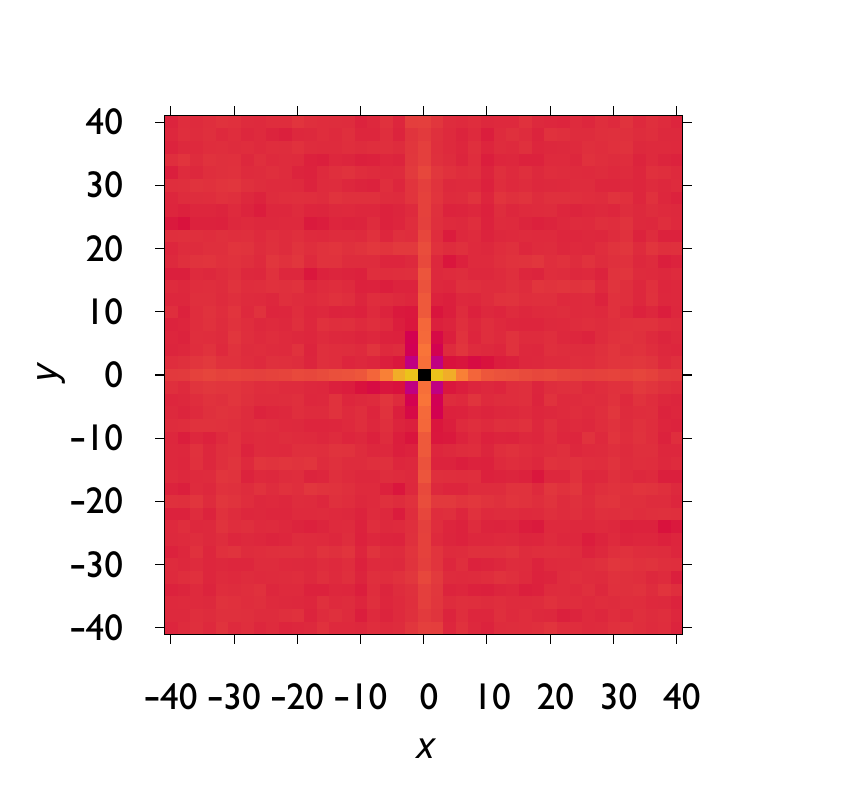}
\includegraphics[angle=0, height=3.1cm, trim=35mm 0 8mm 0]{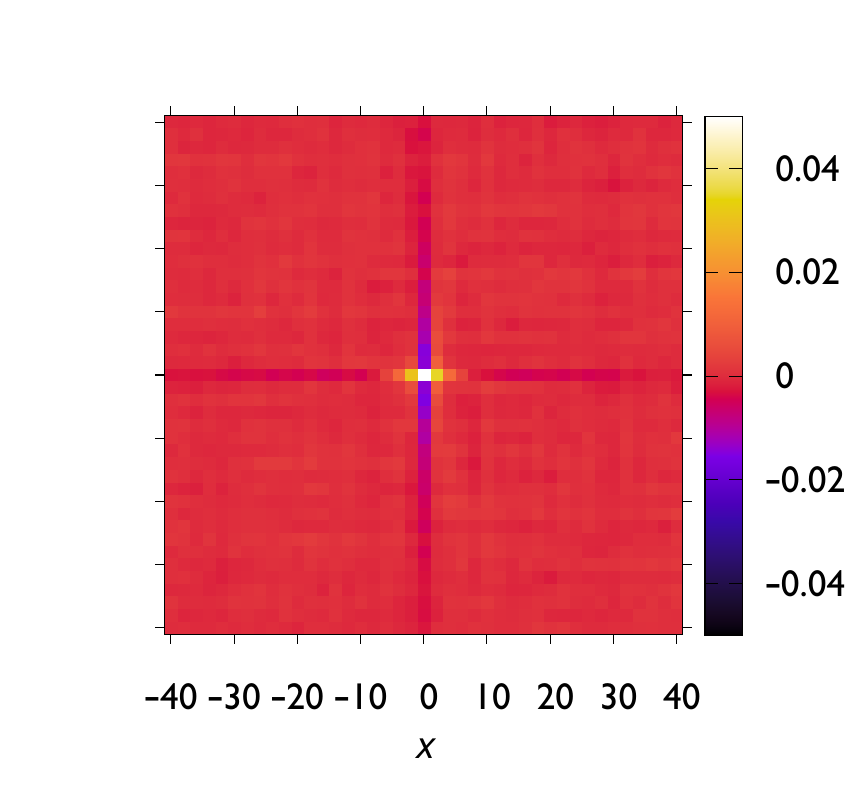}
\vspace*{-2mm}

%
%\hspace*{-10mm}
%\begin{picture}(0,0)
%\put(-12,100){\sffamily{(b)}}
%\end{picture}
%\includegraphics[angle=0, height=4cm, trim=4mm 0 8mm 0]{Figures/stress_strain/stress_strain_comparison.pdf}
\end{center}
\vspace*{-4mm}
\caption{Dislocation pattern evolution in SCDD simulations for different parameters at $\gamma = 1$ and $a=2$. Left two columns: Total and GND density maps obtained at different strain $\gamma$ values. Right two columns: Same sign and opposite sign spatial correlation functions ($d_{++}$ and $d_{+-}$, respectively) of dislocation densities.
\vspace*{-0.7cm}
\label{fig:pattern_evol_continuum_params}}
\end{figure*}

We now address the role of SCDD parameters in the patterning instability. Figure \ref{fig:pattern_evol_continuum_params} plots the dependence of typical patterns and the corresponding correlation functions on the coefficients of the gradient terms $D$ and $A$ and the average strength of the yield threshold $\alpha$. As seen, in the absence of gradient terms ($D=A=0$) the $+-+-$-type instability discussed above can be clearly observed (second row). This behavior is, in agreement with the theory, induced by the friction stress term $\tau_f$ and is at odds with the patterns of DDD and demonstrates the necessity of the inclusion of gradient terms. The third row of Fig.~\ref{fig:pattern_evol_continuum_params} shows the effect of $D<0$: the increased strength of the GND gradient depicted in Fig.~\ref{fig:sketch} leads to even stronger $+-+-$ instability, again in agreement with the linear stability analysis of Ref.~\cite{groma_dislocation_2016}. As of the role of the other parameters, we first recall that $\tau_d$ is a diffusive term in the total dislocation density $\rho$, so increasing the value of $A$ leads to smoothening of the dislocation patterns (see the 4th row of Fig.~\ref{fig:pattern_evol_continuum_params}) while $A<0$ would lead to anti-diffusion and the immediate blow-up of the pattern (not shown). The effect of decreased yield strength is seen in the last row of Fig.~\ref{fig:pattern_evol_continuum_params}: one observes weaker patterns and polarization and the increase of $\alpha$ would lead to the strengthening of the $+-+-$ instability. To summarize, it is evident that in order to obtain realistic patterns $D>0$ and $A>0$ is required and both too large and too small values of $\alpha$ should be avoided. A more detailed analysis will be published elsewhere.

%\begin{figure}
%\includegraphics[angle=0, width=6cm]{Figures/order_parameter/phase_diagram.pdf}
%\caption{The value of parameter $o$ at yielding for $Y=0.5$ and $a=2$ at plastic strain of $\gamma=2$. Significant wall formation can only be observed for $D^*\gtrsim 0.2$.
%\vspace*{-5mm}
%\label{fig:phase_diagram}}
%\end{figure}

It is instructive to compare the continuum plasticity theory with general elastoplastic constitutive models and in particular, with those of kinematic hardening~\cite{Asaro:2006lh}. The backstress term appearing therein has the phenomenological role of modeling the Bauschinger effect observed at reversed loading with the appropriate translation of the yield surface. In this paper, we showed that there is an explicit correspondence: since  $\tau_b$ in Eqs.~(\ref{eq:rhod2plus},\ref{eq:rhod2minus}) can also be considered as an asymmetric correction to $\tau_f$, and using the identity $\partial_t \kappa(\bm r, t) = - \partial_x \dot{\gamma}(\bm r, t)$, with the GND density connected to the shear component $\dot{\gamma}$ of the plastic strain rate, one arrives at $\dot{\tau}_b = (D/\rho) \partial^2_x \dot{\gamma}$ that is analogous to the phenomenological rate equation of Melan and Prager ($\dot{\tau_b} \propto \dot{\gamma}$) \cite{melan_1938, prager_1961} (the appearance of the second derivative reflects the strain-gradient origin of the Bauschinger effect and the backstress in microscopically derived continuum theories of dislocation behavior). The simulations presented above, therefore, emphasize the microscopic origin of the backstress: The asymmetry of the yield surface in kinematic hardening is the result of the bulid-up of asymmetric dislocation sub-structures (polarized walls in the present set-up).
Furthermore, backstress terms are also used in gradient plasticity theories to account for the short-range interactions in pile-ups close to grain boundaries \cite{shizawa_thermodynamical_1999}. Such terms exhibit the same form as Eq.~(\ref{eq:gradient_stress}) also with a positive dimensionless prefactor. 

%The difference can be readily explained by noticing that pile-ups mainly consist of GNDs. In such a case, if the pile-up contains only, e.g., $+$ dislocations (that is, $\kappa = \rho$) the flow-stress disappears from Eq.~(\ref{eq:rhod2plus}) and only the $\rho \ln(\rho)$-type term gives a stress contribution from Eq.~(\ref{eq:plastic_potential}). This means that the two gradient terms merge into one as $\tau_b + \tau_d = -(A/\rho)\partial_x \rho$. Since negative $A$ would lead to anti-diffusion and immediate blow-up of the dislocation structure, it follows that in the fully polarized situation, in accordance with gradient plasticity theories, only one back-stress term remains preceded with a positive prefactor even with $D<0$.

In the case of SCDD the picture that emerges is as follows: the DDWs formed during plastic deformation are in fact two GND pile-ups piling up against each other. The $+-$ instability at the center is due to the flow stress which also provides the great strength of such structures. However, backstress terms are also required to suppress the $+-+-$ instability discussed in the paper and to provide a length-scale for the pile-up widths.

In summary, SCDD does not only provide precise description of its microscopic DDD counterpart, thus representing a successful multi-scale step, but it also synthesizes previous theoretical approaches of dislocation pattern formation, kinematic hardening, and strain gradient plasticity in a simple 2D setting. By identifying the physical interpretation and key role of the strain gradient terms our results may serve as a starting point for more complex 3D implementations.

We thank Michael Zaiser for fruitful discussions. This work has been supported by the National Research, Development and Innovation Office of Hungary (PDI and IG, project Nos.\ NKFIH-K-119561 and NKFIH-KH-125380), the Czech Science Foundation (PDI, project No.\ 15-10821S) and by the U.S. Department of Energy, Office of Sciences, Basic Energy Sciences, DE-SC0014109 (SP). This work was completed in the ELTE Institutional Excellence Program (1783-3/2018/FEKUTSRAT) supported by the Hungarian Ministry of Human Capacities. PDI is also supported by the \'UNKP-18-4 New National Excellence Program of the Hungarian Ministry of Human Capacities and by the J\'anos Bolyai Scholarship of the Hungarian Academy of Sciences.

\bibliography{plasticity}

\end{document}

%%Title.  Two "affiliation" class options are groupedaddress (default) and superscriptaddress
%\title{Supplementary Material -- The Emergence and Role of Dipolar Dislocation Patterns in Discrete and Continuum Formulations}
%\date{\today}
%
%\author{P\'eter Dus\'an Isp\'anovity}
%\affiliation{Department of Materials Physics, E\"otv\"os University, P\'azm\'any P\'eter s\'et\'any 1/a, H-1117 Budapest, Hungary}
%
%\author{Stefanos Papanikolaou}
%\affiliation{The Johns Hopkins University, Hopkins Extreme Materials Institute, Baltimore, MD 21218}
%
%\author{Istv\'an Groma}
%\affiliation{Department of Materials Physics, E\"otv\"os University, P\'azm\'any P\'eter s\'et\'any 1/a, H-1117 Budapest, Hungary}

%\affiliation{The Johns Hopkins University, Hopkins Extreme Materials Institute, Baltimore, MD 21218}

%\maketitle

%\section{Supplementary Information}

\newpage

\onecolumngrid
%\appendix

\begin{center} \textbf{ \large Supplemental Material}
\end{center}

\section{Numerical methods}

\subsubsection{Discrete dislocation dynamics}

The motion of $N$ straight parallel edge dislocations is simulated in single slip. The Cartesian coordinate system is chosen such that the dislocations are parallel to the $z$ axis and the slip direction as well as the Burgers vectors (of equal length $b$) point along the $x$ axes. Possible directions of the Burgers vector $\bm b$ are distinguished by the sign $s=\pm1$ as $\bm b_i = s_i (b,0)$, where $1 \ge i \ge N$. We prescribe dislocation charge neutrality ($\sum_{i=1}^N s_i=0$) and denote the position of the dislocations in the $z=0$ plane as $\bm r_i = (x_i, y_i)$. The simulation area is $L \times L$ in the $xy$ plane and periodic boundary conditions (PBC) are assumed at all sides. Dislocation motion is assumed to be overdamped, so equation of motion of discrete dislocations can be written as:
\begin{equation}
\dot{x}_i(t) = s_i \left[ \tau_\text{ext} + \sum_{j=1, j\ne i}^N s_j \tau_\text{ind}(\bm r_i -\bm r_j) \right]; \ \dot{y}_i(t) = 0,
\label{eq:eq_mot}
\end{equation}
where $\tau_\text{ext}$ is the externally applied shear stress and $\tau_\text{ind}$ denotes the stress field of an individual positive ($s_i=+1$) dislocation. The latter is calculated for PBC by considering an infinite amount of image dislocations both in the $x$ and $y$ directions:
\begin{equation}
\tau_\text{ind}(x,y) = \sum_{i,j=-\infty}^\infty \tau_\text{ind}^\text{ibc}(x-iL,y-jL),
\end{equation}
where
\begin{equation}
\tau_\text{ind}^\text{ibc}(x,y) = \frac{x(x^2-y^2)}{(x^2+y^2)^2}
\end{equation}
is the solution for infinite boundary conditions \cite{hirth_theory_1982}. Note that in the equations above we used the dimensionless units introduced in Table 1 of the main text. The equation of motion (\ref{eq:eq_mot}) is solved by a 4.5th order Runge-Kutta scheme.

The simulations are started from a random (thus non-equilibrium) configuration of dislocations. First, a relaxation step is performed, that is, Eq.~(\ref{eq:eq_mot}) is solved for each dislocation at zero applied stress, then $\tau_\text{ext}$ is increased in a quasistatic manner, allowing relaxation at constant stress every time a strain avalanche sets on (for details see \cite{Ispanovity:2014ff}).

\subsubsection{Stochastic continuum dislocation dynamics}

The evolution of the dislocation densities $\rho$ (total or statistically stores) and $\kappa$ (geometrically necessary, GND) is simulated in 2D on an equidistant grid of $M\times M$ cells each of size $a\times a$. The system size is thus $L \times L = Ma \times Ma$. The evolution of the densities is governed by the phase-field functional
\begin{equation}
P[\rho, \kappa] = E_\text{el} + \int \left[ A \rho \ln\left( \frac{\rho}{\rho_0} \right) + \frac{D}{2} \frac{\kappa^2}{\rho}\right] \mathrm d^2 r,
\label{eq:plastic_potential2}
\end{equation}
where $E_\text{el}$ is the mean-field elastic energy of the system:
\begin{equation}
E_\text{el} = \int \frac 12 \tau_\text{sc}(\bm r) \gamma(\bm r) \mathrm d^2 r.
\end{equation}
According to the theory derived earlier \cite{groma_dislocation_2016}, during the evolution the phase-field potential $P$ must always decrease. This, however, is not of simple steepest descent type, since dislocation flow stress plays a role similar to static friction. This leads to equilibrium configurations that are \emph{not} at (even local) energy minima and also introduce history dependence and frustration into the system.

Previously, flow stress was introduced for this theory as a deterministic function of the local fields: $\tau_f(\bm r) = \alpha \sqrt{\rho(\bm r)}$ in agreement with the Taylor relationship ($\alpha$ is a dimensionless prefactor here)~\cite{Groma:2003pi, groma_variational_2010, groma_dislocation_2016}.   In the current model, however, we rather assume that the flow stress is a stochastic variable in line with previous models of avalanche dynamics both for crystalline~\cite{zaiser_fluctuation_2005} and amorphous matter~\cite{martens_spontaneous_2012}. Its statistics has been recently determined from discrete dislocation dynamics simulations~\cite{ispanovity_role_2017}. It was found that $\tau_f$ follows a Weibull distribution with shape parameter $\beta = 1.4$ and average $\alpha \sqrt{\rho(\bm r)}$:
\begin{equation}
\Phi(\tau_f(\bm r)) = 1 -\exp\left(-\left(\frac{\tau_f(\bm r)}{\tau_0(\bm r)}\right)^\beta \right),
\label{eq:weibull}
\end{equation}
where $\Phi$ denotes the cumulative distribution function and $\tau_0(\bm r) = \alpha \sqrt{\rho(\bm r)} / \Gamma(1+1/\beta)$.

%\renewcommand\thefigure{S\arabic{figure}} 
%
%\begin{figure*}
%\subfigure[]{\includegraphics[angle=0, width=7.5cm]{Figures/D_plus_comparison/rho_D_plus.pdf}}
%\subfigure[]{\includegraphics[angle=0, width=7.5cm]{Figures/D_plus_comparison/kappa_D_plus.pdf}}
%\subfigure[]{\includegraphics[angle=0, width=7.5cm]{Figures/D_plus_comparison/rho_D_minus.pdf}}
%\subfigure[]{\includegraphics[angle=0, width=7.5cm]{Figures/D_plus_comparison/kappa_D_minus.pdf}}
%\caption{\label{fig:polarization} Dislocation density maps for different $D(p)$ functions at $\tau_\text{ext}=0.4$. (a),(b): Total density ($\rho$) and GND density ($\kappa$) for $D(p)=D^*$ and (c),(d): for the $D(p)$ function of the paper for which $D(0) = D^*$ and $D(1) = -D^*$. The figures demonstrate that $D<0$ is necessary for the formation of dipolar dislocation walls of finite length. The simulation parameters are identical to those of Fig.~2 of the main text.}
%\end{figure*}

For the sake of simplicity, in the cellular automaton representation we track the evolution of $\rho_+(\bm r) = [\rho(\bm r) + \kappa(\bm r)]/2$ and $\rho_-(\bm r) = [\rho(\bm r) - \kappa(\bm r)]/2$ signed density fields, since these quantities must always be non-negative. We also identify a dislocation quantum of $\Delta \rho = 1/a^2$, which corresponds to exactly one dislocation (in the dimensionless units used throughout the paper). During the evolution the $\rho_\pm$ densities at every simulation step one dislocation quantum of size $\Delta \rho$ moves from one cell to its neighbor of either $+$ or $-$ sign.

SCDD simulations, in accordance with their DDD counterparts, start from a random pattern of density fields which are generated by assigning a random cell for every positive and negative dislocation quantum (of size $\Delta \rho = 1/a^2$). The total number of these quanta is $M^2*a^2/2$ for both positive and negative sign dislocations since the average total density is unity in the dimensionless units of the paper. Then, a random flow stress value is assigned to each cell according to Eq.~(\ref{eq:weibull}).

As explained in detail in Refs.~\onlinecite{Groma:2003pi} and \onlinecite{groma_variational_2010}, the $1/\rho$ prefactor in the last term of the plastic potential $P$ [Eq.~(\ref{eq:plastic_potential2})] represents the square of a local scale. Since the dislocation systems studied in this paper are scale-free, this scale can only be the local average dislocation spacing $1/\sqrt{\rho}$, in accordance with the principle of similitude (see, e.g.,~Ref.~\onlinecite{zaiser_scaling_2014}). In the current numerical implementation, however, a fixed scale of the cell size is introduced, that is proportional to the average rather than the local dislocation spacing. To account for this fact, the $1/\rho$ prefactor is replaced by $1/\rho_0$ (being equal to unity in the dimensionless units).

%In the main text a certain smooth $D(p)$ function is used in Eq.~(\ref{eq:plastic_potential2}) that fullfills $D(0) = D^*$ and $D(1) = -D^*$. Figure \ref{fig:polarization} demonstrates the necessity of such a crossover function, since in the $D(p)>0$ case no short dipolar walls form, which is characteristic to DDD simulations.

\bibliography{scpm_pattern_v8.6}